\documentclass[preprint,prd,nofootinbib,amsmath,amssymb,superscriptaddress]{revtex4-1}

\usepackage{graphicx}% Include figure files
\usepackage{dcolumn}% Align table columns on decimal point
\usepackage{bm}
\usepackage{enumerate}
\usepackage{hyperref}
\usepackage{xcolor}
\usepackage{mathrsfs}
\usepackage{multirow}
\usepackage{bbold}
\usepackage{ulem}

\allowdisplaybreaks
\newcommand{\nn}{\nonumber}

\newcommand{\Kt}{\vec{K}_{\rm T}}

\newcommand{\Mvev}{\mathcal{M}_{\rm vev}}

\newcommand{\VV}{(V_{\rm CT}+V_{\rm CW})}

\usepackage{xifthen}
\newcommand{\ifequals}[3]{\ifthenelse{\equal{#1}{#2}}{#3}{}}
\newcommand{\CASE}[2]{#1 #2} % Dummy, so \renewcommand has something to overwrite...
\newenvironment{switch}[1]{\renewcommand{\CASE}{\ifequals{#1}}}{}

\newcommand{\dphi}[1]{
    \begin{switch}{#1}
        \CASE{1}{(c_\beta \partial_{\rho_1 } + s_\beta  \partial_{\rho_2 } )}
        \CASE{2}{(c_\beta \partial_{\eta_1 } + s_\beta  \partial_{\eta_2 } )}
        \CASE{3}{(c_\beta \partial_{\zeta_1} + s_\beta  \partial_{\zeta_2} )}
        \CASE{4}{(c_\beta \partial_{\psi_1 } + s_\beta  \partial_{\psi_2 } )}
        \CASE{5}{(c_\beta \partial_{\rho_2 } - s_\beta  \partial_{\rho_1 } )}
        \CASE{6}{(c_\beta \partial_{\eta_2 } - s_\beta  \partial_{\eta_1 } )}
        \CASE{7}{(c_\beta \partial_{\zeta_2} - s_\beta  \partial_{\zeta_1} )}
        \CASE{8}{(c_\beta \partial_{\psi_2 } - s_\beta  \partial_{\psi_1 } )}
    \end{switch}
}

\begin{document}

\title{Global Symmetries and Effective Potential of 2HDM \\ in Orbit Space}

\author{Qing-Hong Cao}
\email{qinghongcao@pku.edu.cn}
\affiliation{School of Physics, Peking University, Beijing 100871, China}
\affiliation{Center for High Energy Physics, Peking University, Beijing 100871, China}

\author{Kun Cheng}
\email{chengkun@pku.edu.cn}
\affiliation{School of Physics, Peking University, Beijing 100871, China}

\author{Changlong Xu}
\email{changlongxu@pku.edu.cn}
\affiliation{School of Physics, Peking University, Beijing 100871, China}

\begin{abstract}
%We extend the framework of analyzing the 2HDM in its orbit space to study the one-loop effective potential before and after electroweak symmetry breaking. In this framework, we present a comprehensive analysis of global symmetries of the one-loop thermal effective potential in the 2HDM, demonstrating when the global symmetries of the tree-level 2HDM potential are broken by loop contributions. By introducing light-cone coordinates and generalizing the bilinear notation around the vacuum, we present a geometric view of the scalar mass matrix and on-shell renormalization conditions. 
We extend the framework of analyzing the 2HDM in its orbit space to study the one-loop effective potential before and after electroweak symmetry breaking. In this framework, we present a comprehensive analysis of global symmetries of the one-loop thermal effective potential in the Higgs family space, demonstrating when the global symmetries in the Higgs family space of the tree-level 2HDM potential are broken by loop contributions. By introducing light-cone coordinates and generalizing the bilinear notation around the vacuum, we present a geometric view of the scalar mass matrix and on-shell renormalization conditions. 
\end{abstract}

\maketitle

\section{Introduction}

The Two-Higgs-Doublet Model (2HDM) is a simple extension of the SM~\cite{Lee:1973iz}. It has received much attention for its potential to provide new sources of CP violation and strong first-order phase transition~\cite{JforScalarFermion,Branco:2005em,Gunion:2005ja,Trautner:2018ipq,Cline:2011mm,Basler:2016obg,Basler:2019iuu,Ferreira:2019bij,Branco:2011iw,THDMbilinear_CPV}.
The most general tree-level 2HDM scalar potential
\begin{equation}\label{eq:V2hdm}
  \begin{split}
    V=&
   m_{11}^2\Phi_1^{\dagger}\Phi_1+m_{22}^2 \Phi_2^{\dagger}\Phi_2
    -\left(m_{12}^2 \Phi_1^{\dagger}\Phi_2 + h.c.\right)  \\
    &
    +\frac{1}{2} \lambda_1 \left(\Phi_1^{\dagger }\Phi_1\right)^2
    +\frac{1}{2} \lambda_2 \left(\Phi_2^{\dagger }\Phi_2\right)^2 +\lambda_3 (\Phi_2^{\dagger}\Phi_2) (\Phi_1^{\dagger }\Phi_1)
    +\lambda_4 (\Phi_1^{\dagger }\Phi_2) (\Phi_2^{\dagger }\Phi_1)    \\
    &+\Big(
    \frac{1}{2} \lambda_5 \left(\Phi_1^{\dagger }\Phi_2\right)^2
    +\lambda _6(\Phi_1^{\dagger }\Phi_1) (\Phi_1^{\dagger }\Phi_2) 
    +\lambda _7 (\Phi_2^{\dagger }\Phi_2) (\Phi_1^{\dagger }\Phi_2) 
    +h.c.    \Big)\\
  \end{split}
\end{equation}
is parameterized by 14 real parameters. Here, ($m_{12}^2,\lambda_5,\lambda_6,\lambda_7 $) are in general complex while the others are real. 

The CP conserving 2HDM, also called real 2HDM, require all the parameters in Eq.~\eqref{eq:V2hdm} to be real with respect to a $U(2)_\Phi$ basis transformation $\Phi_i' = U_{ij}\Phi_j$. Due to the field redefinition, the CP symmetry and other global symmetries of the potential are hard to determine from the parameters in Eq.~\eqref{eq:V2hdm} directly, and one of the most efficient ways to analyze these symmetries is to use the bilinear notation~\cite{THDMbilinear,Ivanov:2006yq,Ivanov:2007de,Nishi:2006tg,Nagel:2004sw,Ivanov:2005hg} of the 2HDM. This method involves expressing the tree-level 2HDM potential in terms of orbits of $SU(2)_L$ gauge transformations, which can be combined to form a four-vector,
\begin{equation}\label{eq:Kcomponents}
(K_0,\vec{K})=K^\mu=\Phi_i^\dagger \sigma_{ij}^\mu \Phi_j,\quad
(\mu=0,1,2,3).
\end{equation}
In this notation, the $U(2)_\Phi$ basis transformation of the Higgs doublets corresponds to a $SO(3)_K$ rotation of the three space-like components of this four-vector, while CP transformations correspond to improper rotations in these three dimensions~\cite{THDMbilinear_CPV}.

The bilinear notation serves as a convenient tool for examining the symmetries and vacuum conditions of 2HDM. However, its applications are usually restricted to tree-level potential and global structures.
In this work, we establish a complete framework for discussing the 2HDM potential, by extending the bilinear notation of the 2HDM to address the properties of physical fields around the vacuum and one-loop effective potential including renormalization.
Recently, it is shown in Ref.~\cite{Cao:2022rgh,Sartore:2022sxh} that the bilinear notation can be extended to Yukawa couplings, making it possible to express the 2HDM effective potential including fermion loop contributions in the bilinear notation~\cite{Cao:2022rgh}. With this approach, we express the effective potential entirely as a function of gauge orbits, and systematically analyze the possible global symmetries of the effective potential. We generalize the bilinear notation to discuss physical fields after electroweak symmetry breaking (EWSB), and provide a geometrical description of scalar masses based on the light-cone coordinates in the orbit space. We demonstrate that the scalar mass matrix can be viewed as a Hessian matrix between two hyper-surfaces in the orbit space. Additionally, we translate the renormalization conditions in the field space into a set of geometrical conditions in the orbit space. Then numerous redundant renormalization conditions~\cite{Basler:2018cwe} that depend on the selection of background fields can be avoided. After the on-shell renormalization, we give a comprehensive effective field theory description of one-loop 2HDM effective potential for the first time.

In the rest of this paper, we first review the global symmetries of the tree-level 2HDM potential in the bilinear notation in Section~\ref{sec:introduce-bilinear-symmetry}, and then examine whether these symmetries are preserved by one-loop corrections in Section~\ref{sec:one-loop-calculation}. We explore the relationship between the orbit space and the field space around the vacuum after EWSB in Section~\ref{sec:bilinear-EWSB}, and we demonstrate how to carry out the on-shell renormalization in the orbit space in Section~\ref{sec:onshell-renormalization}. Finally, we conclude in Section~\ref{sec:conclusion}.

\section{Basis invariant description of global symmetry}\label{sec:introduce-bilinear-symmetry}
The basis and CP transformations of the Higgs doublets and the global symmetries of 2HDM potential in the bilinear notation are originally introduced in  Refs.~\cite{THDMbilinear,Ivanov:2006yq,Ivanov:2007de,Nishi:2006tg,Nagel:2004sw,Ivanov:2005hg}. Our work is based on their framework, and we give a brief introduction in this section.
 If a 2HDM potential is invariant under some basis or CP transformations, then it possesses the corresponding symmetries. 
The bilinear notation is convenient to discuss these global transformations, because the basis or CP transformations simply corresponds to proper or improper rotations in the 3-dimensional space-like part of the orbit space~\cite{Branco:2011iw,THDMbilinear_CPV,Ivanov:2006yq,Ivanov:2007de}, and we refer this 3-dimensional subspace as $\vec{K}$ space in the following.

\subsection{Global transformations and symmetries in the bilinear notation}

We first consider the $U(2)_\Phi$ basis transformations $\Phi_i\to U_{ij}\Phi_j$. It is straightforward to see from Eq.~\eqref{eq:Kcomponents} that an $SU(2)_\Phi$ basis transformation corresponds to a rotation in the $\vec{K}$-space,
\begin{equation}\label{eq:SO3}
  K_0\to K_0,~ K_{a}\to R_{ab}(U)K_b,\quad R_{ab}(U)= \frac{1}{2} \text{tr}\left[ U^\dagger \sigma_a U \sigma_b \right], ~a,b=1,2,3.
\end{equation}

Then we consider the CP transformation $\Phi_i(t,\vec{x})\to \Phi_i^*(t,-\vec{x})$. Because the definition of the standard CP transformation $\Phi_i\to\Phi_i^*$ will be changed if we choose another set of basis to describe the scalar fields, e.g. $\Phi_i' = U_{ij}\Phi_j$, the CP transformations in the 2HDM are extended as~\cite{Ecker:1987qp,Lavoura:1994fv,JforScalarFermion,Gunion:2005ja}
\begin{align}\label{eq:GCP}
    {\rm GCP:}\quad \Phi_i \to X_{ij}\Phi_j^*.
\end{align}
Here, $X_{ij}$ is an arbitrary unitary matrix, and such CP transformations are called generalized CP (GCP) transformations. By plugging the GCP transformation into Eq.~\eqref{eq:Kcomponents}, we find that $\vec{K}$ transforms in an improper $O(3)_K$ rotation $\bar{R}(X)$.
\begin{equation}\label{eq:O3}
  K_0\to K_0,~ K_{a}\to\bar{R}_{ab}(X)K_b,\quad \bar{R}(X)\equiv R(X)\operatorname{diag}(1,-1,1).
\end{equation}
Here, the $R(X)$ is defined in Eq.~\eqref{eq:SO3}. Besides, for any GCP transformation, one can always find a basis $\Phi_i$ so that $X_{ij}$ is a real rotation matrix~\cite{Ecker:1987qp}. Therefore GCP transformations are often classified into three cases~\cite{Ferreira:2009wh,Ferreira:2010yh}:
\begin{align}
\label{eq:CP1}
    {\rm CP1:}&~ \Phi_1 \to \Phi_1^*,~ \Phi_2 \to \Phi_2^*,  \\
\label{eq:CP2}
    {\rm CP2:}&~ \Phi_1 \to \Phi_2^*,~ \Phi_2 \to -\Phi_1^*, \\
\label{eq:CP3}
    {\rm CP3:}&~
    \begin{cases}
      \Phi_1 \to \Phi_1^*\cos{\theta} + \Phi_2^*\sin{\theta} \\
      \Phi_2 \to -\Phi_1^*\sin{\theta} + \Phi_2^*\cos{\theta}
    \end{cases},\quad
    0<\theta<\pi/2,
\end{align}
where CP3 is the most general CP transformation while CP1 and CP2 are some special case with ${\rm CP1}^2={\rm CP2}^2=\mathbb{1}$ up to a global sign. 

After showing that the basis and GCP transformations correspond to $O(3)_K$ rotations in the $\vec{K}$-space, we examine the symmetries conserving conditions on the 2HDM potential.
The 2HDM potential in Eq.~\eqref{eq:V2hdm} can be written as a function of gauge orbits~\cite{THDMbilinear,Ivanov:2006yq,Ivanov:2007de,Nishi:2006tg},
\begin{equation}\label{eq:Vbilinear}
    \begin{split}
        V&=\xi_{\mu} K^{\mu}+ \eta_{\mu \nu} K^{\mu} K^{\nu} \\
        &=\xi_0 K_0 + \eta_{00} K_0^2+\vec \xi \cdot \vec K
        + 2K_0 \vec \eta \cdot \vec K+ \vec K^T E\vec K.
    \end{split}
\end{equation}
Here, $\vec{\xi}$ parametrizes the scalar quadratic couplings while $E$ and $\vec{\eta}$ parametrize the scalar quartic couplings. As discussed above, a GCP or basis transformation corresponds to some (improper) rotation $R$ in the $\vec{K}$-space. If a tree-level potential is invariant under a rotation $R$, i.e., $V(K_0,\vec{K})  = V(K_0,R\vec{K})$, its parameters should be invariant under the rotation $R$,
\begin{equation}\label{eq:treelevelsymmetry}
    \vec{\xi} = R\vec \xi, \quad \vec{\eta}= R\vec{\eta}, \quad
    E = RER^T.
\end{equation}

A complete analysis of the symmetries of the 2HDM Lagrangian must include the Yukawa interaction, which reads as
\begin{equation}
    -\mathcal{L}_{\rm Yuk} = \bar{Q}_L y_{u,i}\tilde{\Phi}_i u_{R}+\bar{Q}_L y_{d,i}\Phi_i d_{R} + h.c.~,\quad i=1,2,
\end{equation}
for one generation of $u$-quark and $d$-quark. 
In this work we concentrate on the Higgs family transformation and assume that the quark fields do not transform under the Higgs family transformation. Under this assumption, we can simplify our discussion to one generation of quarks.
In addition, $y_{d,i}$ and $y_{u,i}^*$ transform like $\Phi_i$ under the $SU(2)_\Phi$ basis redefinition. Although the Yukawa coupling terms in the Lagrangian cannot be expressed in bilinear notation directly, it is shown that the bilinear notation can still be extended to discuss whether the Yukawa couplings break the global symmetries of the scalar potential~\cite{Cao:2022rgh}.
This is done by projecting the Yukawa couplings into orbit space, and defining covariant vectors in the dual space of $K^\mu$ in terms of the Yukawa couplings as
\begin{equation}\label{eq:YukawaY}
 Y_u^\mu=y_{u,i}^* \sigma^\mu_{ij} y_{u,j},\quad
 Y_d^\mu=y_{d,i} \sigma^\mu_{ij} y_{d,j}^*.
\end{equation}
In order to make sure that $\mathcal{L}_{\rm Yuk}$ is invariant under the basis transformation $\Phi_i \to U_{ij}\Phi_j$ or the CP transformation in the scalar sector $\Phi_i \to X_{ij}\Phi_j^*$, the vector $\vec{Y}$ projected by the Yukawa couplings should satisfy
\begin{equation}\label{eq:yukawaSymmetric}
\vec{Y}  =  R(U)\vec{Y} \quad {\rm or}\quad \vec{Y}  =  \bar{R}(X)\vec{Y}.
\end{equation}

\subsection{Examples of global symmetries}

Next we show how to discuss some special symmetries that is widely considered in the orbit space. We start with two characteristic examples, the CP1 symmetry and the $Z_2$ symmetry.
The CP symmetry is often introduced to the potential because large CP violations are prohibited by experiments. From Eq.~\eqref{eq:O3}, the CP1 transformation $\Phi_i\to \Phi_i^*$ corresponds to a mirror reflection in the $\vec{K}$-space.
The $Z_2$ symmetry is introduced to prevent the flavor changing neutral interactions,\footnote{For different types of $Z_2$ charge assignments in the orbit space, see Table~\ref{tab:z2charge} in Appendix~\ref{sub:softz2}.} and a softly broken $Z_2$ symmetry is often considered. From Eq.~\ref{eq:SO3}, the $Z_2$ transformation $\Phi_1\to-\Phi_1$ corresponds to a 2-dimensional rotation of $\pi$ in the $\vec{K}$ space.

\begin{figure}
    \centering
    \includegraphics[width=.3\linewidth]{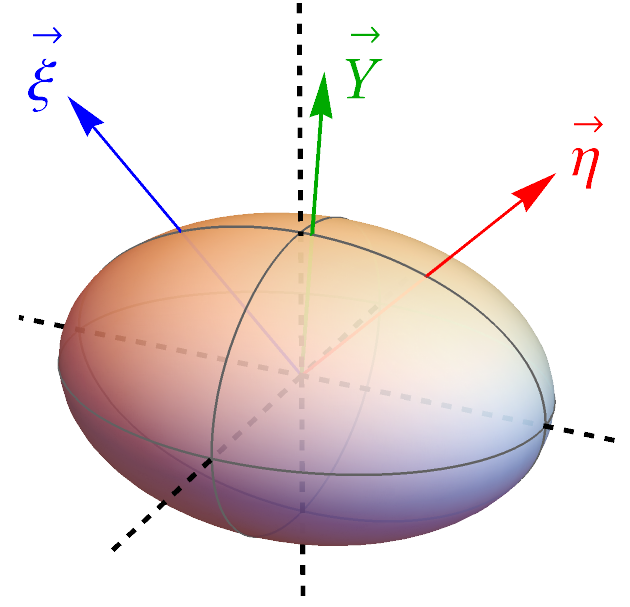}\qquad
    \includegraphics[width=.3\linewidth]{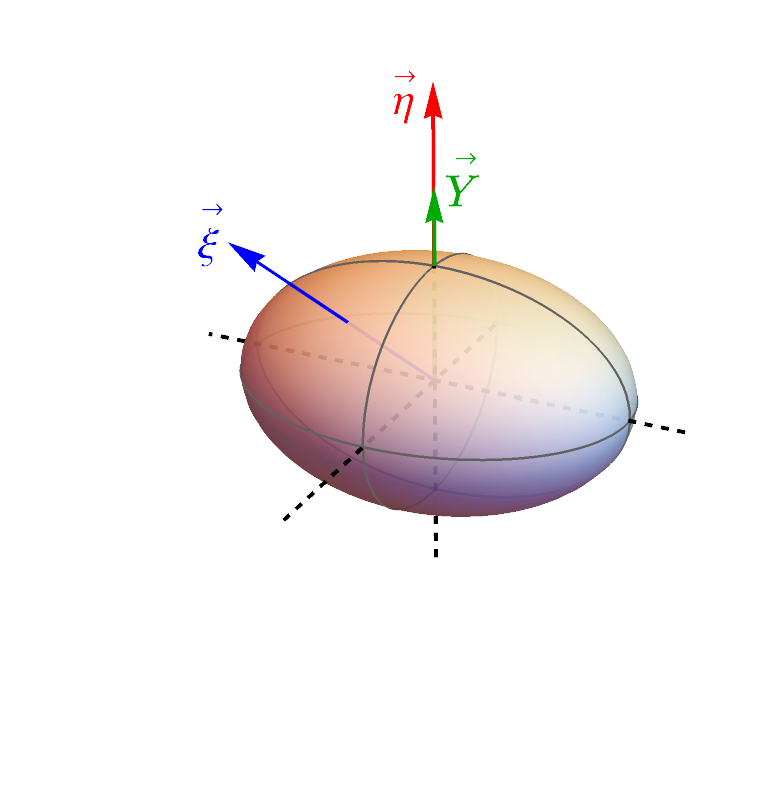}
    \caption{Illustration figure of parameter vectors and tensor of the 2HDM potential that obeys CP1 symmetry (left) or softly broken $Z_2$ symmetry (right). The ellipsoid denotes the tensor $E$ and black dashed lines denote its three principal axes. Red and blue arrows denote the directions of $\vec{\eta}$ and $\vec{\xi}$ respectively.}
    \label{fig:geometricview}
\end{figure}

Whether a 2HDM is invariant under the CP1 or $Z_2$ transformation can be understood from the geometrical profile of parameter tensor and vectors, as shown in Eqs.~\eqref{eq:treelevelsymmetry} and~\eqref{eq:yukawaSymmetric}.
Without loss of generality, we use an ellipsoid to visualize the  $3\times 3$ real symmetric tensor $E$ which possesses at least three $C_2$ axis (principal axis) and three symmetry planes, and we illustrate these two examples in Fig.~\ref{fig:geometricview}.
The CP1 symmetric 2HDM potential satisfies a mirror reflection symmetry in the $\vec{K}$ space, requiring all the parameter vectors to lie on the same reflection plane of $E$. The $Z_2$ symmetric potential is invariant under a rotation of $\pi$ in the $\vec{K}$ space. Hence the parameter vectors should point to the same principal axis of $E$. As for the softly broken $Z_2$ symmetry, the quadratic term $\vec{\xi}$ is allowed to break the $Z_2$ symmetry as in Fig.~\ref{fig:geometricview}.

Following Refs.~\cite{Ferreira:2009wh,Ferreira:2010yh}, we list other global symmetries in scalar family space by different geometric profile of the scalar potential in Table~\ref{tab:profile}. The $U(1)_a$ transformation $\Phi_1\to e^{i\theta}\Phi_1$ corresponds to a rotation along a certain axis, the CP2 transformation corresponds to a point reflection, and the CP3 transformation corresponds to a point reflection followed by an additional rotation of $0\sim \pi$.  The geometric profiles show the hierarchy chain of those global symmetries clearly, 
\begin{equation}
 {\rm CP1} < Z_2 <
    \left\{
    \begin{matrix}
    {\rm CP2} \\
    U(1)_a
    \end{matrix} \right\}
   < {\rm CP3} <  U(2),
\end{equation}
i.e., a $Z_2$ symmetric tree-level 2HDM scalar potential must satisfy CP1 symmetry and likewise. For GCP properties of tree-level 2HDM scalar potential, CP2 and CP3 symmetric conditions are more strict than CP1.\footnote{This situation is different in $N$-Higgs doublet model for $N>2$. A potential that does not conserve CP1 symmetry may satisfy some higher order GCP symmetries~\cite{Ivanov:2015mwl,Ivanov:2018ime}. For detailed discussion of high order GCP symmetries, see Ref.~\cite{Ivanov:2017zjq,Ivanov:2018qni,Ivanov:2018ime,Ivanov:2019kyh}}
Besides, neither CP2 nor CP3 symmetry can be still preserved after the Higgs field developed a non-vanishing vacuum expectation value. Therefore, we will only discuss CP1 conserving (CPC) conditions and denote CP1 as CP in the following.

\begin{table}[h]
    \centering
    \caption{Global symmetry and their geometric profile of parameter vectors and tensor $E$. $\vec{e}_i$ are direction of three eigenvectors of $E$, and $e_i$ denotes the three corresponding eigenvalues.  Note that the vector $\vec{Y}$ should satisfy the same constrain with $\vec{\eta}$ and $\vec{\xi}$ if we assume that the quark fields do not transform under Higgs family transformation.} 
    \begin{tabular}{|c|c|c|c|c|}
        \hline  Symmetry & Transformation & Vector $\vec{\eta}$ and $\vec{\xi}$ & Tensor  \\
        \hline  U(2) &$\Phi_i \to U_{ij}\Phi_j$& 0 & spherical \\
        \hline  CP3  & Eq.~(8) & 0 & $e_1=e_2$ \\
        \hline  CP2  & Eq.~(7) & 0 & - \\
        \hline  $U(1)_a$ &$\Phi_1 \to e^{i\theta}\Phi_1$& collinear with $\vec{e}_3$ & $e_1=e_2$ \\
        \hline $Z_2$ &$\Phi_{1,2} \to \pm\Phi_{1,2}$& collinear with an axes $\vec{e}_i$ & - \\
        \hline  CP1  &$\Phi_i \to \Phi_i^*$& orthogonal to an axes $\vec{e}_i$ & - \\
        \hline
    \end{tabular}
    \label{tab:profile}
\end{table}

At the end of this section, we would like to mention that additional symmetry-related opportunities will appear when we include three generations and allow the quark fields to transform together with the scalar fields.
In that case, $y_{d,i}$ and $y_{u,i}^*$ do not simply transform like $\Phi_i$, and their transformation rule is determined by how the quark fields transform together with the scalar fields.
For example, one can construct CP2 symmetric model with non-trival Yukawa sector~\cite{Maniatis:2007de}. Under the CP2 transformation, instead of transforming as a point reflection $\vec{Y}\to -\vec{Y}$, the vector $\vec{Y}$ transform between different families such as $\vec{Y}^m\leftrightarrow -\vec{Y}^n$, where $m$ and $n$ are family indexes.
In addition, the symmetries in Table~1 is a summary of the symmetries in the Higgs family transformation only. The 2HDM Lagrangian can have many more symmetries after considering non-trival transformations in the fermion family. For example, in the $Z_3$ symmetric 2HDM~\cite{Ferreira:2010ir,Ivanov:2013bka}, the scalar potential is invariant under the $U(1)$ symmetry, but the Yukawa couplings are only invariant under $Z_3$ symmetry, breaking the symmetry of the potential to a finite subgroup of $U(1)$.
Instead of exploring all symmetry-related aspects of the 2HDM Lagrangian, in this work we focus on the transformation and symmetries in the Higgs family space. Unless otherwise stated, the word ``symmetry" in this paper only refers to the symmetry in the Higgs family transformation.

\section{Effective potential and thermal correction}\label{sec:one-loop-calculation}

The global symmetries of thermal effective potential are important in the study of the vacuum structure and CP violation in the early universe. The use of bilinear notation simplifies, from a geometrical perspective, the analysis of the global symmetries of the effective potential~\cite{Ivanov:2008er,Ginzburg:2009dp}. In this section, we employ the bilinear notation to evaluate the complete one-loop effective potential, and discuss its global symmetries.

The thermal effective potential of the 2HDM is written as
\begin{equation}\label{eq:VeffT}
    V_{\rm eff}(T)=V_{\rm tree}+V_{\rm CW} + V_T + V_{\rm daisy} ,
\end{equation}
where $V_{\rm tree}$ is the tree-level potential, $V_{\rm CW}$ is the one-loop Coleman-Weinberg potential at zero temperature~\cite{Coleman:1973jx}, and $V_T+V_{\rm daisy}$ are the thermal corrections at finite temperature. Using the background field method, the one-loop Coleman-Weinberg potential calculated in Landau gauge under the $\overline{\rm MS}$ scheme is
\begin{equation}
    \begin{aligned}
       V_{\rm CW}(\phi_c) &=\frac{1}{2}\mathbf{Tr}\int \frac{d^4p_E}{2\pi^4}\ln \left[p^2_E+\mathbf{M}^2(\phi_c)\right]\\
       &= \frac{1}{64\pi^2} \sum_i n_i m_i^4(\phi_c) \left[ \ln \frac{m_i^2(\phi_c)}{\mu^2} - c_i \right].
       \label{eq:Vcw}
 \end{aligned}
\end{equation}
Here, $p_E=\left(-i p^0, \vec{p}\right)$, $\mathbf{M}^2$ is the mass matrix of scalar or fermion in the loop and $\mathbf{Tr}$ traces over the dimension of mass matrix, $m_i^2$ is the eigenvalue of the $\mathbf{M}^2$ for the field $i$, and $n_i$ is the degree of freedom of the field $i$. The constant $c_i$ equals to 5/6 for gauge bosons and 3/2 for others.

The effective potential of the 2HDM has been extensively studied in the literature~\cite{Cline:2011mm,Basler:2019iuu,Basler:2016obg,Ferreira:2019bij,Bernon:2017jgv}. Typically, only the neutral or CP-even components of the Higgs boson doublets are treated as background fields, which breaks the $SU(2)_L$ invariance explicitly. Consequently, the bilinear notation cannot be applied to study $V_{\rm eff}(\phi_c)$ directly. In order to analyze the global symmetries of the effective potential using bilinear notation, a global $SU(2)_L$ invariance must be preserved in the calculation~\cite{Cao:2022rgh}, which means that the masses in Eq.~\eqref{eq:Vcw} need to be evaluated in a $SU(2)_L$ invariant way. To achieve this, we treat all the components of the Higgs boson doublets $\Phi_i$'s,
\begin{equation}
\Phi_i=\begin{pmatrix}
  \phi_{i\uparrow}\\\phi_{i\downarrow}
\end{pmatrix}, \quad i=1,2,
\end{equation}
as background fields, and $K^\mu$ should be understood as bilinear forms of background fields in this section. 

\subsection{Symmetries of Coleman-Weinberg potential}
We first consider the zero temperature effective potential by calculating the contributions from gauge boson loop, fermion loop and scalar loop to the Coleman-Weinberg potential respectively.

\paragraph{Contributions from gauge boson loop.}
The masses of gauge bosons arise from the kinetic term $|D_\mu\Phi_i|^2$ with $D^\mu \Phi_i = (\partial^\mu + i\frac{g}{2} \sigma_{a} W_{a}^\mu + i\frac{g'}{2}B^\mu)\Phi_i$, where $i=1,2$.
Expanding the covariant derivative term directly yields the gauge boson mass term
\begin{equation}
    \begin{split}
        \frac{1}{4}\Phi_i^\dagger(g\sigma_a W_a + g' B)^2\Phi_i =&
        \frac{1}{4}\Phi_i^\dagger(g'^2 B^2 + 2gg' B W_a \sigma_a + g^2 \sigma_a \sigma_b W_a W_b )\Phi_i  \\
        =&\frac{1}{4}\Phi_i^\dagger(g'^2 B^2 + 2gg' B W_a \sigma_a + g^2 \sigma_{\{a} \sigma_{b\}} W_a W_b )\Phi_i  \\
        =&\frac{\Phi_i^\dagger\Phi_i}{4}(g'^2 B^2 + g^2 W_a W_a) + \frac{\Phi_i^\dagger \sigma_a \Phi_i}{2} gg'B W_a.
    \end{split}
\end{equation}
Then the gauge boson mass matrix in basis $\vec{G}=(W_1,W_2,W_3,B)$ is 
\begin{equation}\label{eq:GaugeBosonMass}
    \mathbf{M}^2_G(\Phi_i)= \frac{\partial^2 L}{\partial \vec{G} \partial \vec{G}}
    =\frac{g^2}{4}\begin{pmatrix}
        \Phi_i^\dagger\Phi_i&0&0&t_W\Phi_i^\dagger\sigma_1\Phi_i \\
        0&\Phi_i^\dagger\Phi_i&0&t_W\Phi_i^\dagger\sigma_2\Phi_i \\
        0&0&\Phi_i^\dagger\Phi_i&t_W\Phi_i^\dagger\sigma_3\Phi_i \\
        t_W\Phi_i^\dagger\sigma_1\Phi_i&t_W\Phi_i^\dagger\sigma_2\Phi_i&t_W\Phi_i^\dagger\sigma_3\Phi_i&t_W^2\Phi_i^\dagger\Phi_i
    \end{pmatrix}
\end{equation}
where $t_W=\tan\theta_W=g'/g$. For a matrix with the shape of Eq.~\eqref{eq:GaugeBosonMass}, its eigenvalues are     
\begin{equation}
    \text{Eigenvalues}
    \begin{pmatrix}
     e &&&a\\
     &e&&b\\
     &&e&c\\
     a&b&c&d
    \end{pmatrix}
    =(e,e,\frac{d+e\pm \sqrt{4(a^2+b^2+c^2)+(d-e)^2}}{2}).
\end{equation}
With the help of the Fierz identities,
\begin{align}
(\Phi_i^\dagger\sigma_a\Phi_i)(\Phi_j^\dagger\sigma_a\Phi_j)&=(\Phi_1^\dagger\Phi_1-\Phi_2^\dagger\Phi_2)^2+4(\Phi_1^\dagger\Phi_2)(\Phi_2^\dagger\Phi_1)= |\vec K|^2,
\end{align}
we present four eigenvalues of the gauge boson mass matrix,
\begin{equation}\label{eq:WZmass}
\begin{aligned}
    m^2_{W^\pm}&=\frac{g^2}{4}K_0,\\
    m^2_Z&=\frac{g^2}{8}\left( (1+t_W^2)K_0 + \sqrt{4t^2_W|\vec K|^2+(t_W^2-1)^2K_0^2} \right),\\
    m^2_\gamma&=\frac{g^2}{8}\left( (1+t_W^2)K_0 - \sqrt{4t^2_W|\vec K|^2+(t_W^2-1)^2K_0^2} \right).
\end{aligned}
\end{equation}
Notice that there is a massless photon when the vacuum is neutral, i.e., $K_0 = |\vec K|$. By plugging Eq.~\eqref{eq:WZmass} into Eq.~\eqref{eq:Vcw}, we find that the gauge boson loop contributions to the Coleman-Weinberg potential, $V_{\rm CW}^{(G)}=V_{\rm CW}^{(G)}(K_0,|\vec K|)$, is spherically symmetric and preserve any rotational symmetry in the $\vec{K}$ space, i.e.,
\begin{equation*}
    V_{\rm CW}^{(G)}(K_0,\vec{K})= V_{\rm CW}^{(G)}(K_0,R\vec{K}),\qquad R\in O(3).
\end{equation*}

\paragraph{Contributions from the quark loop.}
Typically, only the contribution from the heaviest quark needs to be included in the effective potential. However, we include both the top and bottom quarks in our calculation to ensure an explicit $SU(2)_L$ invariance. The top and bottom quark masses mix due to the presence of charged background fields, and the fermion mass matrix given by $-\partial^2\mathcal{L}/\partial \bar\psi^i_{L}\partial \psi^j_{R} $ is
\begin{equation}
    (\bar{t}_L,  \bar{b}_L)
    \mathbf{M}_F
    \begin{pmatrix}
     t_R\\
     b_R
    \end{pmatrix},\quad
    \mathbf{M}_F=
    \begin{pmatrix}
     y_{it}\phi_{i\downarrow}^* & y_{ib}\phi_{i\uparrow} \\
     -y_{it}\phi_{i\uparrow}^* & y_{ib}\phi_{i\downarrow}
    \end{pmatrix}.
\end{equation}
We obtain the fermion masses after singular decomposition,
\begin{equation}\label{eq:tbmass}
    L^{-1}\mathbf{M}_F R =
    \begin{pmatrix}
    m_t & \\ & m_b
    \end{pmatrix},
    \quad
    m_{t/b}^2= \dfrac{B \pm \sqrt{B^2+C}}{2},
\end{equation}
where, with the help of vector $\vec{Y}$ defined in Eq.~\eqref{eq:YukawaY}, $B$ and $C$ can be written as $SO(3)_K$ basis invariant forms as follows:
\begin{equation}\label{eq:BC}
    \begin{split}
    B=&\frac{1}{2}(Y_{t0} + Y_{b0} )K_0+ \frac{1}{2}(\vec Y_t + \vec Y_b )\cdot \vec{K}, \\
    C=&-\frac{1}{2}(Y_t\cdot Y_b)K_0^2
    - K_0(Y_{t0}\vec{Y}_b+Y_{b0}\vec{Y}_t)\cdot\vec{K}\\
    & +\frac{1}{2}\vec{K}\cdot(\vec Y_t\cdot\vec Y_b-Y_{t0}Y_{b0}-\vec Y_t\otimes\vec Y_b-\vec Y_b\otimes\vec Y_t)\cdot\vec{K}.
    \end{split}
\end{equation}
The masses can be simplified in the case that the Yukawa couplings exhibit a large hierarchy; for example, when $y_t\gg y_b$, only the top quark mass $m_t^2(K)=(Y_{t0}K_0+\vec Y_t\cdot\vec K)/4$ needs to be considered. Equations~\eqref{eq:tbmass} and \eqref{eq:BC} show that the symmetry of $V_{\rm CW}^{(F)}$ is completely determined by the direction of vector $\vec{Y}$. When the vector $\vec{Y}$ is invariant under the rotation, i.e., $\vec{Y}_{t/b} = R \vec{Y}_{t/b}$ for $R\in O(3)$,
\begin{equation*}
    V_{\rm CW}^{(F)}(K_0,\vec{K})= V_{\rm CW}^{(F)}(K_0,R\vec{K}).
\end{equation*}
When $\vec{Y}_{t/b} \neq R \vec{Y}_{t/b}$ ,
\begin{equation*}
    V_{\rm CW}^{(F)}(K_0,\vec{K})\neq V_{\rm CW}^{(F)}(K_0,R\vec{K}).
\end{equation*}
Therefore, whether the fermion loop contribution to $V_{\rm CW}$ breaks the global symmetry of the tree-level potential depends on the pattern of Yukawa couplings.

\paragraph{Contributions from the scalar loop.}
The calculation of $V^{(S)}_{\rm CW}(K_0,\vec{K})$ can be performed straightforwardly from Eq.~\eqref{eq:Vcw}, in which the mass matrix of scalars is given by
\begin{equation}
    \mathbf{M}_S^2(\varphi)_{ab}=\frac{\delta^2V}{\delta\varphi_{a}\delta\varphi_{b}},
\end{equation}
where $\varphi_a$ are real vectors in the 8-dimensional field space.
Though $\mathbf{M}_S^2$ cannot be diagonalized analytically, we still find a way to investigate the global symmetries of $V^{(G)}_{\rm CW}$. We firstly employ the notations in Ref.~\cite{Degee:2009vp}, where the components of $\varphi_a$ are ordered as
\begin{equation}
    \varphi_{a}^{T}=\left(\operatorname{Re} \phi_{1 \uparrow}, \operatorname{Im} \phi_{1 \uparrow}, \operatorname{Re} \phi_{2 \uparrow}, \operatorname{Im} \phi_{2 \uparrow}, \operatorname{Re} \phi_{1 \downarrow}, \operatorname{Im} \phi_{1 \downarrow}, \operatorname{Re} \phi_{2 \downarrow}, \operatorname{Im} \phi_{2 \downarrow}\right),
\end{equation}
and $\varphi_a$ is related to the bilinear form by $K^{\mu}=\varphi_{a} \Sigma_{a b}^{\mu} \varphi_{b}$. The $8\times8$ matrices $\Sigma^\mu$ are defined as
\begin{equation}
\label{sigma}
    \Sigma^\mu=\Sigma^\mu_4\oplus\Sigma^\mu_4,\quad
    \Sigma_4^0=\mathbb{1}_4,\quad
    \Sigma_4^1=\begin{pmatrix}
     0 & \mathbb{1}_2\\
     \mathbb{1}_2 & 0\\
    \end{pmatrix},\quad
    \Sigma_4^2=\begin{pmatrix}
     0 & \mathbb{i}_2\\
     -\mathbb{i}_2 & 0\\
    \end{pmatrix},\quad
    \Sigma_4^3=\begin{pmatrix}
     \mathbb{1}_2 & 0\\
     0 & -\mathbb{1}_2\\
    \end{pmatrix},
\end{equation}
where $\mathbb{1}_d$ is the $d\times d$ identity matrix and $\mathbb{i}_2\equiv\left(\begin{smallmatrix} 0&1\\-1&0 \end{smallmatrix}\right).$

The $V_{\rm CW}^{(S)}$ can be expanded in the powers of $\mathbf{M}_S$~\cite{Quiros:1999jp},
\begin{equation}
    \begin{aligned}
       V_{\rm CW}^{(S)} &=\frac{1}{2}\mathbf{Tr}\int \frac{d^4p_E}{2\pi^4}\ln \left[p_E^2+\mathbf{M}^2_S\right]\\
      &=\frac{1}{2}\int
\frac{d^4p_E}{2\pi^4}\left[\mathbf{Tr}\sum_{n=1}^\infty\frac{1}{n}\left(-\frac{\mathbf{M}_S^2}{p^2_E}\right)^n+\ln p^2_E\right],
       \label{eq:Vcws}
 \end{aligned}
\end{equation}
where $\mathbf{Tr}$ stands for taking a trace over the 8-dimensional field space. For example, the leading power is
\begin{equation}\label{eq:scalarmass}
\mathbf{Tr}(\mathbf{M}^{2}_{S})
    =\left(20\eta_{00}+4\operatorname{tr}(E)\right)K_0+24\vec{K}\cdot\vec{\eta}+8\xi_0,
\end{equation}
which is consistent with Ref.~\cite{Degee:2009vp}. We show that all the traces  $\mathbf{Tr}(\mathbf{M}_S^{2n})$ in Eq.~\eqref{eq:Vcws} are functions of gauge orbits $K^\mu$, and the complete calculations are deferred to Appendix~\ref{appsub:scalar}. Here, we present the final calculation result expressed in the bilinear notation as
\begin{equation}\label{eq:Vs}
        V_{\rm CW}^{(S)} = \mathcal{F}\left(S_{p}^{\mu\nu}, \eta^{\mu\nu} \right) .
\end{equation}
The function $\mathcal{F}$ only depends on the trace of the inner products of $S_{p}^{\mu\nu}$ and $\eta^{\mu\nu}$, and $S_p^{\mu\nu}$ is defined as 
\begin{equation}
    S_p^{\mu\nu}=F(p)^{\mu}K^{\nu}+F(p)^{\nu}K^{\mu}-g^{\mu\nu}(F(p)K).
\end{equation}
Here, $F(p)_\mu$ is a function of $K^\mu$ that depends on the integer $p$,
\begin{align}
 F(p)_0 &\equiv
 \sum_{k=0}^{p/2} C_p^{2k} (A_0)^{p-2k} |\vec{A}|^{2k},\\
 \vec{F}(p) &\equiv\left\{
 \begin{aligned}
 - &\sum_{k=0}^{(p-1)/2} C_p^{2k+1} (A_0)^{p-2k-1}|\vec{A}|^{2k} \vec{A}\qquad &(p\neq0),\\
 &0\qquad &(p=0),
 \end{aligned}
 \right.
\end{align}
where $(A_0,\vec A)=A_{\mu}=2\eta_{\mu\nu}K^{\nu}+\xi_{\mu}$ and $C_p^k$ is the binomial coefficient. Notice that the global symmetries are determined only by the 3-dimension vector $ \vec{A}$.

Upon expressing $V_{\rm CW}^{(S)}$ as a function in the orbit space, we find that the tensor structures in $V_{\rm CW}^{(S)}$ are constructed entirely by tree level parameter tensors $\eta^{\mu\nu}$ and $\xi^\mu$, i.e., no new tensor structure appears, therefore, the rotation symmetries of $V_{\rm CW}^{(S)}$ in the $\vec{K}$-space are determined by the tree-level parameter tensors. If the tree-level potential is invariant under a rotation $R\in O(3)$ in the $\vec{K}$-space, i.e., $V_{\rm tree}(K_0,\vec{K})= V_{\rm tree}(K_0,R\vec{K})$, the scalar loop contribution $V_{\rm CW}^{(S)}$ also preserves the rotation invariance,
\begin{equation*}
    V_{\rm CW}^{(S)}(K_0,\vec{K})= V_{\rm CW}^{(S)}(K_0,R\vec{K}).
\end{equation*}

\subsection{Symmetries of thermal potential}
As for the finite temperature corrections in Eq.~\eqref{eq:VeffT}, $V_T$ stands for the contribution from one-loop diagrams, and $V_{\rm daisy}$ denotes the correction from higher loop Daisy diagrams~\cite{Quiros:1999jp}. The one-loop correction $V_T$ is given by
\begin{equation}
V_{T}=\sum_i n_i\frac{T^4}{2\pi^2}J_{B/F}\left(m_{i}^{2} / T^{2}\right),
\end{equation}
where the thermal bosonic function $J_{B}$ and fermionic function $J_{F}$ are
\begin{align}
J_{B}\left(m^{2} / T^{2}\right)=&-\frac{\pi^{4}}{45}+\frac{\pi^{2}}{12} \frac{m^{2}}{T^{2}}-\frac{\pi}{6}\left(\frac{m^{2}}{T^{2}}\right)^{3 / 2}-\frac{1}{32} \frac{m^{4}}{T^{4}} \log \frac{m^{2}}{a_{b} T^{2}} \nn\\
&-2 \pi^{7 / 2} \sum_{\ell=1}^{\infty}(-1)^{\ell} \frac{\zeta(2 \ell+1)}{(\ell+1) !} \Gamma\left(\ell+\frac{1}{2}\right)\left(\frac{m^{2}}{4 \pi^{2} T^{2}}\right)^{\ell+2},\\
J_{F}\left(m^{2} / T^{2}\right)=& \frac{7 \pi^{4}}{360}-\frac{\pi^{2}}{24} \frac{m^{2}}{T^{2}}-\frac{1}{32} \frac{m^{4}}{T^{4}} \log \frac{m^{2}}{a_{f} T^{2}} \nn\\
&-\frac{\pi^{7 / 2}}{4} \sum_{\ell=1}^{\infty}(-1)^{\ell} \frac{\zeta(2 \ell+1)}{(\ell+1) !}\left(1-2^{-2 \ell-1}\right) \Gamma\left(\ell+\frac{1}{2}\right)\left(\frac{m^{2}}{\pi^{2} T^{2}}\right)^{\ell+2}.
\end{align}
Here, $a_b=16a_f=16\pi^2 e^{3/2-2\gamma_E}$, and $\zeta$ is the Riemann-$\zeta$ function. The leading $T$-dependent terms of $J_{B/F}$ are given by the mass-square terms, 
\begin{equation}
J_{B}= \frac{\pi^{2}}{12} \frac{m^{2}}{T^{2}} + O(T^{-4}),\quad 
J_{F}=- \frac{\pi^{2}}{24} \frac{m^{2}}{T^{2}}+ O(T^{-4}),
\end{equation}
where the background-field-independent terms are dropped. By collecting the results in Eqs.~\eqref{eq:WZmass}, \eqref{eq:tbmass} and \eqref{eq:scalarmass}, we obtain the leading contributions from gauge boson loops, fermion loops, and scalar loops to $V_T$ as follows:
\begin{align}
    \label{eq:VTG}
    V_{T}^{(G)}&\approx \frac{g^2 T^2}{32}(3+t_W^2)K_0,\\
    \label{eq:VTF}
    V_{T}^{(F)}&\approx  -\frac{T^2}{8}\left[(Y_{t0}+Y_{b0})K_0+(\vec{Y}_t+\vec{Y}_b)\cdot\vec K) \right],\\
    \label{eq:VTS}
    V_{T}^{(S)} &\approx \frac{T^2}{6}\left[\left(5\eta_{00}+\operatorname{tr}(E)\right)K_0+6\vec{K}\cdot\vec{\eta}+2\xi_0\right].
\end{align}
We find that the corrections from Eqs.~\eqref{eq:VTG}-\eqref{eq:VTS} to the tree-level potential is equivalent to shifting the quadratic couplings $\xi_\mu$ in the orbit space, i.e., 
\begin{align}\label{eq:shiftxi}
\xi_0&\to\xi_0+T^2 c_{T0}, \nn \\
\vec{\xi}&\to\vec{\xi}+T^2 \vec{c}_{T}, 
\end{align}
where
\begin{align}
 c_{T0}&=\frac{g^2}{32}(3+t_W^2)-\frac{Y_{t0}+Y_{b0}}{8}+\frac{5 \eta_{00}+\operatorname{tr}(E)}{6}, \nn \\
   \label{eq:vecxiT}
    \vec{c}_T&=\frac{1}{8}\left[8\vec \eta-\vec{Y}_t-\vec{Y}_b\right].
\end{align}
The direction of $\vec{\xi}$ is shifted by the quartic couplings $\vec{\eta}$ and Yukawa interactions $\vec{Y}$ from thermal corrections. At a sufficient high temperature with $T^2\gg |\vec{\xi}|/|\vec{c}_T|$, the direction of shifted $\vec\xi$ is aligned along the direction of $\vec c_{T}$. As a result, the symmetries of thermal effective potential under the basis transformation and CP transformation are determined by $\vec c_{T}$.

At high temperatures, the contribution from higher loop Daisy diagrams $V_{\rm daisy}$ is comparable with $V_T$, and it is given by~\cite{Quiros:1999jp}
\begin{equation}
    V_{\rm daisy}= - \frac{T}{12\pi}\sum_{i=\rm bosons}n_i \left[ \mathcal{M}_i^3(\phi_c,T) - m_i^3(\phi_c) \right].
\end{equation}
Here, the $\mathcal{M}_i(\phi_c,T)$ are thermal corrected masses calculated from $V_{\rm tree}+V_T$, which is obtained from the tree level potential by parameter shifting $\xi^\mu \to \xi^\mu + T^2 c_T^\mu$. Therefore, the $T$-dependent terms in $\mathcal{M}_i(\phi_c,T)$ are in the form of $T^2c_T^\mu$. As $c_{T}^0$ plays no role in global transformations, the behavior of $V_{\rm daisy}$ under the $O(3)_K$ transformation depends only on $\vec{c}_T$.

After understanding the behavior of $V_{\rm CW}$, $V_{T}$ and $V_{\rm daisy}$ under the $O(3)_K$ transformation, we are ready to discuss whether a global symmetry preserved by the tree-level potential will be violated by the loop corrections. Consider a tree-level potential that processes the symmetry of a basis or CP transformation, then the potential is invariant under a rotation $R$ in the $\vec{K}$-space, $V_{\rm tree}(K_0,\vec{K})= V_{\rm tree}(K_0,R\vec{K})$, and its parameters satisfy
\begin{equation}\label{eq:treelevelsymmetry2}
    \vec{\xi} = R\vec \xi, \quad \vec{\eta}= R\vec{\eta}, \quad
    E = RER^T.
\end{equation}
The {\it only} quantum correction that may violate the symmetry is the contribution from fermion loops. The global symmetry is maintained in effective potential if and only if all the Yukawa couplings are invariant under $R$, i.e., $\vec{Y}= R\vec{Y}$.

If the symmetry is softly broken at tree level, i.e., only the scalar quadratic coupling $\vec{\xi}\neq R\vec{\xi}$ violates the symmetry while other conditions in Eq.~\eqref{eq:treelevelsymmetry2} are preserved,
then the symmetry violation effect from soft terms tend to be suppressed at high temperature.
This is because the leading thermal corrections shift the scalar quadratic couplings $\vec{\xi}$ with Yukawa coupling $\vec{Y}$ and scalar quartic couplings $\vec{\eta}$, and both $\vec{Y}$ and $\vec\eta$ preserve the symmetry.

Another noteworthy example is the custodial symmetry. In the orbit space, the custodial symmetry of the 2HDM does not correspond to a rotation symmetry but a shift symmetry~\cite{Grzadkowski:2010dj}. As the effective potential is not invariant under any shift symmetry in the orbit space, the custodial symmetry of 2HDM is bound to be broken by the effective potential.

\section{Bilinear notation after EWSB}\label{sec:bilinear-EWSB}

In this section, we extend the bilinear notation to discuss EWSB and physical fields. There are two reasons to discuss the EWSB in the bilinear notation.

\begin{figure}
    \centering
        \includegraphics[width=0.3\linewidth]{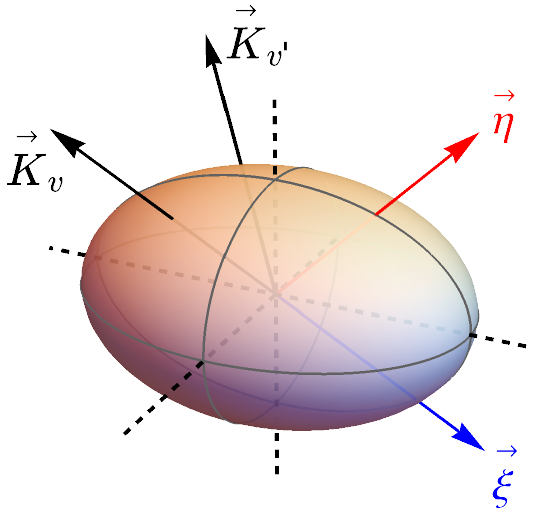}
\caption{Illustration figures of tree-level parameters for an SCPV potential. Here $\vec K_v$ and $\vec K_{v'}$ are a pair of degenerated vacuum expectation values that are related by a mirror reflection, and the reflection plane spanned by $\vec{\xi}$ and $\vec{\eta}$. 
    }
    \label{fig:ellipsoidvev}
\end{figure}

Firstly, a global symmetry exhibited by the potential, as shown in Table~\ref{tab:profile}, can be broken spontaneously after the potential develops a vacuum.  For example, consider the CP symmetry. Even if the potential is explicitly CP conserving, $V(\Phi_1,\Phi_2)|_{\Phi_i\to\Phi_i^*}=V(\Phi_1,\Phi_2) $, the physical fields, which are fluctuations around the vacuum, may still break CP symmetry after EWSB if the vacuum has an  additional CP phase as follows:
\begin{equation}\label{eq:vphase}
    \langle \Phi_1 \rangle = \begin{pmatrix} 0\\v_1  \end{pmatrix},\quad
    \langle \Phi_2 \rangle = \begin{pmatrix} 0\\v_2e^{i\delta}  \end{pmatrix}.
\end{equation}
This is called spontaneous CP violation (SCPV). In the bilinear notation, the SCPV happens when the potential but not the vacuum is invariant under a mirror reflection. In this case, there are two degenerate vacua related by a CP transformation as in Fig.~\ref{fig:ellipsoidvev}. 
After analyzing the vacuum conditions in the orbit space, we can easily determine whether the CP symmetry or other global symmetry is spontaneously broken.

Secondly, exploring the physical fields after the EWSB is necessary for performing on-shell renormalization. The renormalized effective potential can be expressed fully in the bilinear notation if we can perform the on-shell renormalization in the orbit space. For that, we examine the vacuum structures in the orbit space and investigate the relations between the field space and orbit space. Furthermore, we demonstrate that the mass matrix of the physical neutral scalars corresponds to a geometric structure in the orbit space, making it convenient to handle the mass spectrum and on-shell renormalization.

\subsection{Vacuum condition}
\begin{figure}
    \centering
    \includegraphics[width=.35\linewidth]{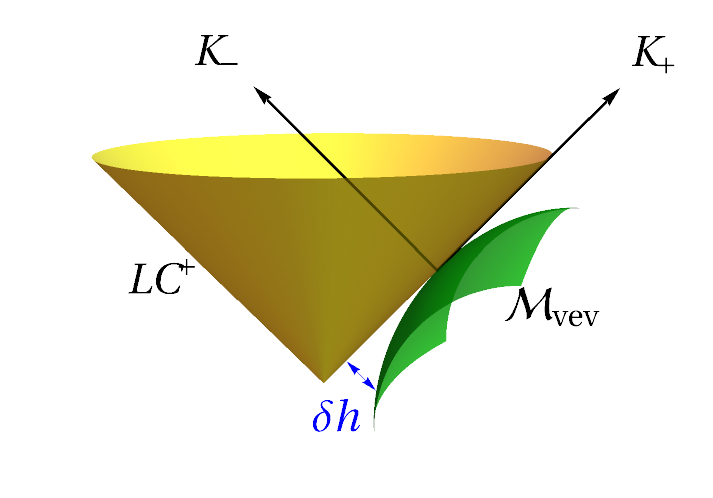}
\caption{Vacuum expectation value and light-cone coordinates in the orbit space. The yellow surface denotes $LC^+$ and the green denotes the equipotential surface $\Mvev$. $K_v$ is the tangent point of these 3-dimensional hyper-surfaces.}
\label{fig:lightcone}
\end{figure}

We start with the vacuum conditions of $V(K^\mu)$, where $V(K^\mu)$ represents the tree-level or effective potential in the orbit space. 
Figure~\ref{fig:lightcone} displays the light-cone in the orbit space, and the light-cone is a hyper-surface defined by $K_0=|\vec{K}|$.
The orbit space inside the forward light-cone $LC^+$ is the physical region, satisfying $K_0\geq|\vec{K}|$~\cite{THDMbilinear,Ivanov:2006yq,Ivanov:2007de,Nishi:2006tg}. A neutral vacuum expectation value requires the minimum of the potential, denoted as $K_v^\mu$, to lie on the $LC^+$, i.e., $K_{v,0}=|\vec{K}_v|$~\cite{THDMbilinear,Ivanov:2006yq,Ivanov:2007de,Nishi:2006tg}. Therefore, $K_v^\mu$ is a conditional minimum of $V(K^\mu)$ on the $LC^+$.

The vacuum of the potential $V(K^\mu)$ is solved by minimizing the function $V_u(K^\mu)=V(K^\mu)-uL(K^\mu)$, where $u$ is a Lagrange multiplier and $L(K^\mu)=0$ is the light-cone condition with $L(K^\mu)$ defined as
\begin{equation}
L(K_0,\vec{K})=K_0^2-|\vec{K}|^2=4K_+K_--|\vec{K}_T|^2.
\end{equation}
Instead of solving the vacuum conditions in the original coordinates chosen in Ref.~\cite{THDMbilinear,Ivanov:2006yq,Ivanov:2007de,Nishi:2006tg}, we introduce the light-cone coordinates
\begin{equation}
K_{\pm}\equiv\frac{(K_0\pm K_3)}{2},\qquad\Kt\equiv(K_1,K_2),
\end{equation}
which are defined after rotating the vacuum along the $K_3$ direction, i.e., $ K_v^\mu = \frac{v^2}{2}(1,0,0,1)^T $.
We will soon see that the geometrical structures of 2HDM vacuum are exposed straightforwardly in the light-cone coordinates.
The solution of the conditional minimum satisfies
\begin{align}\label{eq:minimalcondition}
  \frac{\partial V}{\partial K_-}\bigg|_{K_v}=2v^2u>0,\quad\frac{\partial V}{\partial K_+}\bigg|_{K_v}=0,\quad  \frac{\partial V}{\partial \Kt}\bigg|_{K_v}=0.
\end{align}
Note that we require $\frac{\partial V}{\partial K_-}>0$ to ensure no global minimum inside the light-cone to avoid a charged vacuum. 

In addition to the conditions in Eq.~\eqref{eq:minimalcondition}, we need to make sure that $K_v$ is a minimal point rather than a saddle point. 
In the 4-dimensional orbit space, $K_v$ is the tangent point of $LC^+$ and an equipotential surface $\Mvev$ defined by $V(K^\mu)=V(K^\mu_v)$~\cite{Ivanov:2007de}, and the normal direction of their tangent space is $K_-$, as shown in Fig.~\ref{fig:lightcone}. Therefore, the requirement that $K_v$ is not a saddle point indicates that $\Mvev$ must be outside of the $LC^+$.
Equivalently, if we expand the infinitely small deviation $\delta h$ between $LC^+$ and $\Mvev$ at their tangent point, (as shown in Fig.~\ref{fig:lightcone})
\begin{align}\label{eq:positiveM33}
    \delta h &= (\delta K_+, \delta \Kt)~ \mathbf{M}_{\rm dev}^2
    \begin{pmatrix}
      \delta K_+\\
      \delta \Kt
    \end{pmatrix},\quad \\
    \mathbf{M}_{\rm dev}^2&=\frac{1}{\partial V/\partial K_-}
    \begin{pmatrix}
     \frac{\partial^2 V_u}{\partial K_+^2 } & \frac{\partial^2 V_u}{\partial K_+ \partial \Kt}\\
     \frac{\partial^2 V_u}{\partial K_+ \partial \Kt} & \frac{\partial^2 V_u}{\partial \Kt^2}
    \end{pmatrix}= \frac{1}{\partial V/\partial K_-} \mathbf{H}_{V_u}, \nonumber
\end{align}
the hessian matrix $\mathbf{M}_{\rm dev}^2$ must be positive definite. Note that the hessian matrix of the deviation between the two surfaces is simply proportional to the Hessian matrix of $V_u$, $\mathbf{H}_{V_u}$. % the matrix $H_{V_u}$ Hessian matrix is multiplied by a normalize factor $1/(\partial V/\partial K_-)$, ensuring that the deviation $\delta h$ has the same dimension as $K$.
As to be shown later, the Hessian matrix of the deviation between the two hyper-surfaces directly yields the neutral scalar mass matrix.

Now we have introduced the vacuum conditions fully in the orbit space. These conditions apply to both the tree-level and the effective potentials. Specifically, the tree-level potential in
Eq.~\eqref{eq:Vbilinear} can be written
in terms of the light-cone coordinates as follows,
\begin{align}
    V_{\rm tree}= & \xi_+ K_+ + \xi_- K_- + \vec{\xi}_{\rm T}\cdot \Kt + (K_+,K_-,\Kt)
    \begin{pmatrix}
        \eta_{++} & \eta_{+-} & \vec{\eta}_{\rm T+}\\
        \eta_{+-} & \eta_{--} & \vec{\eta}_{\rm T-}\\
        \vec{\eta}_{\rm T+} & \vec{\eta}_{\rm T-} & {\eta}_{\rm TT}
    \end{pmatrix}
    \begin{pmatrix}
     K_+\\
     K_-\\
     \Kt
    \end{pmatrix}.
\end{align}
Then the minimal conditions for the tree-level potential from Eq.~\eqref{eq:minimalcondition} are
\begin{align}
  &\frac{\partial V_{\rm tree}}{\partial K_-}\bigg|_{K_v}=\xi_- + v^2 \eta_{+-}=2v^2u>0, \nn\\
  & \frac{\partial V_{\rm tree}}{\partial K_+}\bigg|_{K_v}=\xi_+ +  v^2\eta_{++} =0, \nn\\
  & \frac{\partial V_{\rm tree}}{\partial \Kt}\bigg|_{K_v}= \vec{\xi}_{\rm T} + v^2 \vec{\eta}_{\rm T+} =0,
\end{align}
which are equivalent to the minimal conditions given in Ref.~\cite{THDMbilinear}.

\subsection{A geometrical view of the scalar mass matrix}
After the potential develops a vacuum expectation value, the scalar fields become massive.  The field components after the EWSB are fluctuations around the vacuum. Without loss of generality,  we use the Higgs basis in which the vacuum $v$ is rotated to the first doublet, and the field components are
\begin{equation}\label{eq:higgsbasis}
   H_1=\begin{pmatrix}
     G^+\\
     \dfrac{v+\phi+iG^{0}}{\sqrt{2}} 
   \end{pmatrix},~~~~
   H_2=\begin{pmatrix}
     H^+\\
     \dfrac{R+iI}{\sqrt{2}} 
   \end{pmatrix},
\end{equation}
where $\phi,R,I$ and $H^\pm$ are physical fields while $G_0$ and $G^\pm$ are Goldstone fields. By substituting the field components of $H_i$ into Eq.~\eqref{eq:Kcomponents}, and rewriting them in terms of the light-cone coordinates, we have
\begin{equation}\label{eq:expandK}
  \begin{pmatrix}
    K_+\\
    K_1\\
    K_2\\
    K_-
  \end{pmatrix}
  =
  \frac{v^2}{2}\begin{pmatrix}
    1\\
    0\\
    0\\
    0
  \end{pmatrix}+
  v\begin{pmatrix}
    \phi \\
    R \\
    I \\
    0
  \end{pmatrix}+
  \begin{pmatrix}
    \frac{\phi^2}{2}+\frac{G_0^2}{2} + G^+G^- \\
    \phi R + I G_0 + G^+H^- + G^- H^+ \\
    \phi I - R G_0 +i(G^+H^- - G^- H^+) \\
    \frac{I^2}{2}+\frac{R^2}{2}+H^- H^+
  \end{pmatrix}.
\end{equation}

The charged Higgs boson mass is given by
\begin{equation}\label{eq:Mcharged}
m_{H^\pm}^2=
   \left.\frac{\partial V}{\partial H^-H^+}\right|_{\rm vev}=\left.\frac{\partial V}{\partial K_-}\right|_{K_v}\left.\frac{\partial K_-}{\partial H^-H^+}\right|_{\rm vev}=\left.\frac{\partial V}{\partial K_-}\right|_{K_v} .
\end{equation}
As for the neutral physical scalars $\phi,R$ and $I$, their mass matrix is calculated by expanding the potential in the field space as follows,
\begin{equation}\label{eq:Mneutral}
    \delta V = (\delta \phi, \delta R, \delta I)~ \mathbf{M}^2_{\rm neutral}~
    \begin{pmatrix}
     \delta \phi\\ \delta R\\ \delta I
    \end{pmatrix},
\end{equation}
where $\delta \phi,\delta R$ and $\delta I$ are small expansions of the fields around the vacuum. Equation~\eqref{eq:expandK} shows that the three directions $(K_+,\Kt)$, which span the tangent space of $LC^+$ and $\Mvev$, are linearly related to the three neutral scalar fields $(\phi,R,I)$ around the vacuum. The linear relationship between field space and orbit space directly links the scalar mass matrix and the Hessian matrix between $LC^+$ and $\Mvev$.
By combining Eq.~\eqref{eq:Mneutral} with Eqs.~\eqref{eq:positiveM33} and~\eqref{eq:expandK}, we obtain
\begin{equation}\label{eq:MMrelation}
   \mathbf{M}^2_{\rm neutral} =
    v^2\begin{pmatrix}
     \frac{\partial^2 V_u}{\partial K_+^2 } & \frac{\partial^2 V_u}{\partial K_+ \partial \Kt}\\
     \frac{\partial^2 V_u}{\partial K_+ \partial \Kt} & \frac{\partial^2 V_u}{\partial \Kt^2}
    \end{pmatrix}=  v^2 \mathbf{H}_{V_u}.
\end{equation}
Therefore, the neutral mass matrix is simply proportional to the Hessian matrix between the two hyper-surfaces $LC^+$ and $\Mvev$. This geometric picture moves beyond the results in Ref.~\cite{THDMbilinear,Ivanov:2006yq,Ivanov:2007de,Nishi:2006tg}. Treating mass matrix as geometrical structure helps us to simplify the discussion of 2HDM scalar mass spectrum. Below are some examples. 

The experimentally preferred Higgs alignment limit can be read out from Eq.\eqref{eq:MMrelation} directly. In the alignment limit, the neutral scalar $\phi$ in Eq.~\eqref{eq:higgsbasis} corresponds to the SM-like Higgs boson, and all of its properties are very close to the SM Higgs boson, including mass, gauge couplings, Yukawa couplings, and CP property.
Technically, the alignment limit is reached when the neutral scalar $\phi$ in Eq.~\eqref{eq:higgsbasis} is approximately the 125~GeV mass eigenstate and does not mix with other neutral scalars, therefore, we obtain the following relations from Eq.~\eqref{eq:MMrelation},
\begin{equation}\label{eq:alignment}
    \left.\frac{\partial^2 V_u}{\partial K_+ \partial \Kt}\right|_{K_v}=
    \left.\frac{\partial^2 V}{\partial K_+ \partial \Kt}\right|_{K_v}\approx 0,
\end{equation}
where $K_+$ and $\Kt$ are light-cone coordinates in orbit space. At tree-level, this condition yields $\vec{\eta}_{\rm T+}\approx0$ straightforwardly.

\begin{figure}
    \centering
    \includegraphics[width=.4\linewidth]{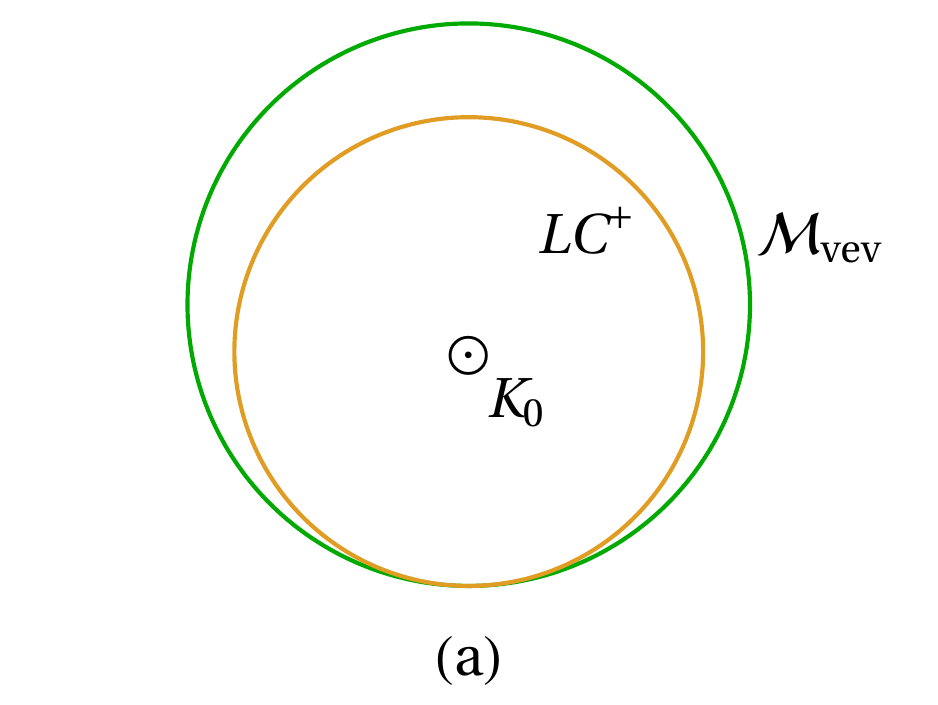}
    \includegraphics[width=.4\linewidth]{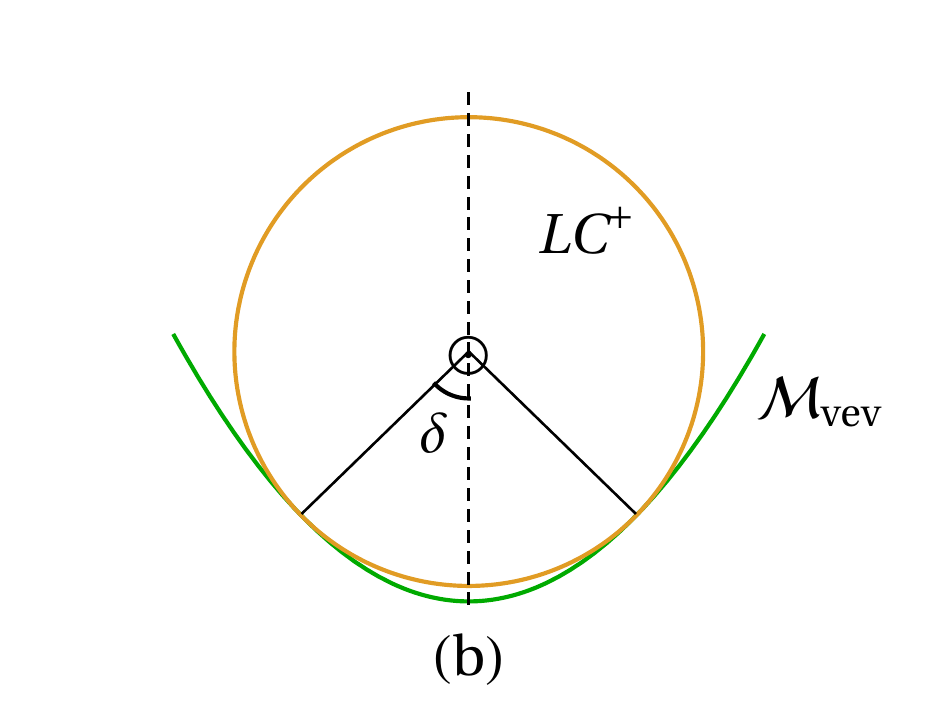}
    \caption{A two dimensional slice of Fig.~\ref{fig:lightcone} with $K_0=K_{v,0}$, viewed from the $K_0$ direction. The symbol $ \odot$ denotes the $K_0$ axis. The yellow line denotes $LC^+$ and the green denotes $\Mvev$. 
    There are two scenarios with an ultra-light scalar: (a) potential with a slightly broken $U(1)_a$ symmetry;
(b) the SCPV potential with a small CP phase $\delta$.  }
    \label{fig:osculation}
\end{figure}

Another demonstration is to discuss the ultra-light CP-odd particle, which is also known as the axion-like particle (ALP). The ALP is of widespread interest for its rich phenomenology, and the 2HDM is a simple model that can provide the ALP. From the geometric relations in the orbit space, a massless scalar appears when the two hyper-surfaces $LC^+$ and $\mathcal{M}_{\rm vev}$ osculate at $K_v$ along a certain direction. 
There are two possibilities in the 2HDM to produce an ALP naturally, due to symmetries rather than accidental parameter choice.
One possibility is the 2HDM potential with an approximately $U(1)_a$ symmetry. An exact $U(1)_a$ symmetry in the 2HDM potential results in an additional Goldstone boson, and the Goldstone boson will develop a small mass if the $U(1)_a$ symmetry is slightly broken as shown in Fig.~\ref{fig:osculation}(a). In this case, the ALP is a pseudo-Goldston boson as in the Dine–Fischler–Srednicki–Zhitnitsky axion model~\cite{Zhitnitsky:1980tq,Dine:1981rt}. Another possibility is the 2HDM potential with a CP symmetry that is spontaneously broken.  When the SCPV phase $\delta$ is very small, the two degenerate vacuums turn to merge, and the two hyper-surfaces $LC^+$ and $\mathcal{M}_{\rm vev}$ turn to osculate with each other at $K_v$, as shown in Fig.~\ref{fig:osculation}(b), therefore, a massless boson appears when the SCPV phase $\delta$ goes to zero~\cite{ZhuSCPV}. In this case, the ALP is not a pseudo-Goldston boson.

\section{On-shell renormalization in the orbit space}\label{sec:onshell-renormalization}
The masses and mixing angles of physical states derived from the one-loop CW potential in the $\overline{\rm MS}$ renormalization scheme differ from their tree-level values. To directly use the loop-corrected masses and mixing angles as inputs, the on-shell renormalization scheme is often preferred. This is achieved by adding the counterterm potential $V_{\rm CT}$ to the zero temperature effective potential
\begin{equation}
    V_{\rm eff}=V_{\rm tree}+V_{\rm CW}+V_{\rm CT},
\end{equation}
and then enforcing the loop-corrected vacuum and masses
to be the same as the tree-level values. Consequently, the renormalization conditions in the field space are given by
\begin{align}
    &\partial_{\varphi_a}(V_{\rm CT}+V_{\rm CW})\big|_{\varphi_a=\langle\varphi_a\rangle_{\rm tree}}=0,\\
    &\partial_{\varphi_a}\partial_{\varphi_b}(V_{\rm CT}+V_{\rm CW})\big|_{\varphi_a=\langle\varphi_a\rangle_{\rm tree}}=0,
\end{align}
where $\varphi_a~(a=1\cdots8)$ denote the eight scalar field components in the two Higgs doublets.

However, most of the renormalization conditions are redundant due to unphysical fields and quite a few identities, and it is convenient to deal with the renormalization condition in orbit space.\footnote{A detailed analysis of the number of renormalization conditions in the field space and their equivalence with the conditions in the orbit space is presented in Appendix~\ref{app:renormalization}.} To achieve this, we express the counterterm potential in the bilinear notation as $V_{\rm CT}=\delta\xi_{\mu}\, K^{\mu}+ \delta\eta_{\mu \nu}\, K^{\mu} K^{\nu} $. Based on the vacuum conditions in Eq.~\eqref{eq:minimalcondition} and the scalar masses given in Eqs.~\eqref{eq:Mcharged} and \eqref{eq:MMrelation}, we obtain ten independent renormalization conditions that are related to the physical fields as follows:
\begin{align}
\label{eq:onshellK1}
    & 0=\partial_{K_+}(V_{\rm CT}+V_{\rm CW})\big|_{K_v},\\
\label{eq:onshellK2}
    & 0=\partial_{\vec K_T}(V_{\rm CT}+V_{\rm CW})\big|_{K_v},\\
\label{eq:onshellK3}
    & 0=\partial_{K_-}(V_{\rm CT}+V_{\rm CW})\big|_{K_v}, \\
\label{eq:onshellK4}
    & 0=\partial_{K_+}^2(V_{\rm CT}+V_{\rm CW})\big|_{K_v}, \\
\label{eq:onshellK5}
    & 0=\partial_{\vec K_T}^2(V_{\rm CT}+V_{\rm CW})\big|_{K_v}, \\
\label{eq:onshellK6}
    & 0=\partial_{K_+}\partial_{\vec K_T}(V_{\rm CT}+V_{\rm CW})\big|_{K_v}.
\end{align}
Here the light-cone coordinates are defined still by the tree-level vacuum $K_v$, and the derivatives are evaluated around $K_v$. Note that only part of the first and second derivatives $\partial_{K^\mu}(V_{\rm CT}+V_{\rm CW})\big|_{K_v}$ and $\partial_{K^\mu K^\nu}(V_{\rm CT}+V_{\rm CW})\big|_{K_v}$ are related to the vacuum conditions and scalar masses and should be included in the renormalization conditions, while the others are irrelevant to physical quantities. Specifically, four conditions from the first derivative in Eqs.~\eqref{eq:onshellK1}-\eqref{eq:onshellK3} ensure that the loop-corrected vacuum expectation value is the same as the tree-level case, and Eq.~\eqref{eq:onshellK3} also ensures that the charged scalar mass is the same as the tree-level value. The other six conditions involving the second derivatives in Eqs.~\eqref{eq:onshellK4}-\eqref{eq:onshellK6} ensure that the neutral scalar masses and mixing angles are the same as those of the tree-level potential.

The counterterms $\delta\xi_\mu$ and $\delta \eta_{\mu\nu}$ can be determined from the renormalization conditions in Eqs.~\eqref{eq:onshellK1}-\eqref{eq:onshellK6}.
For a general 2HDM without any constrains on the parameters, there are fourteen free parameters, four in $\delta \xi_\mu$ and ten in $\delta \eta_{\mu\nu}$, to be determined by the renormalization conditions. After expressing $\delta \xi_\mu$ and $\delta \eta_{\mu\nu}$ in terms of the light-cone coordinates, the renormalization conditions are
\begin{align}
    \label{eq:deltaeta1}
    \delta \eta_{++} &= - \partial_{K_+}^2 V_{\rm CW}\big|_{K_v}, \\
    \label{eq:deltaeta2}
    \delta \vec\eta_{T+} &= - \partial_{\vec K_T}\partial_{\vec K_+} V_{\rm CW}\big|_{K_v}, \\
    \label{eq:deltaeta3}
    \delta \eta_{TT} &= - \partial_{\vec K_T}^2 V_{\rm CW}\big|_{K_v}, \\
    \label{eq:deltaxi1}
    \delta \xi_+ &= - \partial_{K_+} V_{\rm CW}\big|_{K_v}-v^2 \delta\eta_{++}, \\
    \label{eq:deltaxi2}
    \delta \vec \xi_T &= - \partial_{\vec K_T} V_{\rm CW}\big|_{K_v} - v^2 \delta\vec\eta_{T+}, \\
    \label{eq:deltaxi3}
    \delta \xi_- &= - \partial_{K_-} V_{\rm CW}\big|_{K_v}-v^2\delta\eta_{+-}.
\end{align}
Note that neither the vacuum condition nor the scalar mass matrix depends on the counterterms $ \delta \eta_{--} $, $ \delta \eta_{+-} $ and $ \delta\vec\eta_{T-} $, therefore, these four parameters are up to free choices. 

In addition,  our convention is to set the tadpole terms to zero whenever possible.
Generally, one can allow the development of vacuum in the field space and introduce the tadpole terms in $V_{\rm CT}$ as done in Refs.~\cite{Basler:2017uxn,Basler:2018cwe}. However, for the most general 2HDM potential, there will be more parameters than renormalization conditions and we can always set the tadpole terms to zero. Tadpole terms may be necessary if we require the counterterms to satisfy some specific constraints such that the remaining parameters cannot satisfy the renormalization conditions.

For the 2HDM with some specific parameter constraints required by symmetries or alignment, it is a common practice to demand the counterterms $\delta\xi_\mu$ and $\delta\eta_{\mu\nu}$ satisfying the same constraints as the tree-level parameters $\xi_\mu$ and $\eta_{\mu\nu}$. Then the number of parameters in $\delta \xi_\mu$ and $\delta \eta_{\mu\nu}$ is less than fourteen as in the general 2HDM, and the renormalization conditions need to be dealt with case-by-case. For illustration, we discuss the renormalization conditions used in three 2HDMs below.

\paragraph{Softly broken $Z_2$ symmetric potential.}
Imposing a softly broken $Z_2$ symmetry on the 2HDM Lagrangian is the most popular way to prevent flavor-changing neutral interactions. For a complex 2HDM with softly broken $Z_2$ symmetry, the $Z_2$ symmetry gives four additional constraints on $\delta\eta_{\mu\nu}$, and the remaining six counterterms can be fixed by the six conditions in Eqs.~\eqref{eq:deltaeta1}-\eqref{eq:deltaeta3}. The soft quadratic couplings are not constrained, and four parameters in $\delta\xi^{\mu}$ can be fixed by the four conditions in Eqs.~\eqref{eq:deltaxi1}-\eqref{eq:deltaxi3}.

\paragraph{Real 2HDM with softly broken $Z_2$ symmetry.}
In addition to the softly broken $Z_2$ symmetry, a CP symmetry is often imposed on the potential. The tree-level potential is invariant under a mirror reflection $\bar{R}$ in the orbit space, $V_{\rm tree}(K_0, \vec K)=V_{\rm tree}(K_0,\bar{R}\vec K)$.  Note that the CP symmetry does not impose any additional constraint on the quartic counterterms $\delta\eta^{\mu\nu}$, as the $Z_2$ symmetry provides stronger constraints than the CP symmetry. On the other hand, the softly broken terms are constrained by the CP symmetry. Say that the mirror reflection is along the second direction $\bar{R}:K_2\to-K_2$, then $\delta\xi_2$ should be set to zero, leaving three free parameters in $\delta\xi^\mu$. 

Usually, the three parameters in $\delta\xi^\mu$ are not enough to satisfy the four equations in Eqs.~\eqref{eq:deltaxi1}-\eqref{eq:deltaxi3}. But when the vacuum is invariant under the CP transformation, e.g., $K_v^\mu=\frac{v^2}{2}(1,0,0,1)$ and $\vec{K}_v=\bar{R}\vec{K}_v$, there are only three independent conditions in Eqs.~\eqref{eq:deltaxi1}-\eqref{eq:deltaxi3}, because the CW potential satisfies the CP symmetry, $V_{\rm CW}(K_0,\vec{K})=V_{\rm CW}(K_0,\bar{R}\vec{K})$, and we have
\begin{align}\label{eq:dVddVsymmetry}
    \partial_{K_2} V_{\rm CW}\big|_{K_v}=0,\quad \partial_{K_2}\partial_{K^\mu} V_{\rm CW}\big|_{K_v}=0.
\end{align}
Then one renormalization condition $\delta\xi_2=-\partial_{K_2}V_{\rm CW}-v^2\delta\vec{\eta}_{2+}=0$ automatically holds from Eq.~\eqref{eq:deltaxi2}, and we end up with three parameters and three conditions.

However, if the vacuum develops an SCPV phase $\delta$, the CP symmetry is broken spontaneously. The vacuum $\vec{K}_v$ is no longer invariant under the CP transformation, e.g., $K_v^\mu=\frac{v^2}{2}(1,0,\sin\delta,\cos\delta)$ and $\vec{K}_v\neq\bar{R}\vec{K}_v$. As a result, Eqs.~\eqref{eq:dVddVsymmetry} no longer hold. The rest three parameters in $\delta\xi^\mu$ are not enough to satisfy the renormalization conditions if we still require the counterterm $\delta\xi_2=0$. The remaining renormalization condition, which is equivalent to $\partial_{\delta}(V_{\rm CW} +V_{\rm CT})=0$, cannot be fulfilled, and this corresponds to a change of the SCPV phase $\delta$. It could be fixed with a tadpole counterterm of the CP-violating vacuum.

\paragraph{2HDM with the exact alignment.} In the 2HDM, the exact alignment condition requires that the neutral scalar $\phi$ in Eq.~\eqref{eq:higgsbasis} is the 125~GeV mass eigenstate, then the tree-level parameters satisfy $\vec\eta_{T+}=0$ as shown in Eq.~\eqref{eq:alignment}. However, the alignment condition is not protected by any symmetry, and there is no guarantee that the counterterms $\delta\vec\eta_{T+}= - \partial_{\vec K_T}\partial_{\vec K_+} V_{\rm CW}$ vanish. Therefore, the alignment condition is usually broken by quantum corrections.

\section{Conclusion and Discussion}\label{sec:conclusion}

We performed a complete analysis of the CP and basis transformation symmetries of the 2HDM in the orbit space. We extended the study of the global symmetries in orbit space to one-loop thermal effective potential. We demonstrated that the global symmetries of the tree-level potential are preserved by quantum corrections from boson loop contributions, but may be broken by fermion loop contributions, depending on the Yukawa interactions. 

In order to study the vacuum conditions and physical masses in the orbit space, we introduced the light-cone coordinates and generalized the bilinear notation to study the physical scalar fields around the vacuum. It provides a geometric view of the scalar mass matrix and on-shell renormalization conditions. By translating the on-shell renormalization conditions of the vacuum and scalar mass into geometric conditions in the orbit space, we calculated the renormalized one-loop effective potential completely.

We extend our study to the case after the EWSB. The geometrical view of scalar masses can provide insight into special limits of the 2HDM mass spectrum, such as alignment limit and ultra-light scalars, thereby simplifying the analysis. The renormalization conditions are much simpler to be dealt with in the orbit space, and there are at most 10 independent on-shell renormalization conditions for a general 2HDM potential.  Our work provides a foundation for future study of the 2HDM effective potential and its implications in orbit space.

\textbf{Note added:}
During the completion of this paper, some new symmetries of the tree-level 2HDM potential were purposed~\cite{Ferreira:2023dke}, which are based on the transformation $K_0\to -K_0$. Our formalism can also be used to analysis these new symmetries. For example, the gauge loops contribute terms like $g^2 K_0$ to the effective potential, breaking the symmetry $K_0\to-K_0$.

\begin{acknowledgements}
The work is supported in part by the National Science Foundation of China under Grants No. 11725520, No. 11675002 and No. 12235001.
\end{acknowledgements}

\appendix

\section{Basis invariant notations of 2HDM potential}

\subsection{Explicit expression of bilinear notation}
The explicit expression for each component of $K^\mu$ is
\begin{equation}
    K^\mu=\Phi_i^\dagger \sigma_{ij}^\mu \Phi_j=
    \begin{pmatrix}
     \Phi_1^\dagger\Phi_1+\Phi_2^\dagger\Phi_2 \\
     \Phi_1^\dagger\Phi_2+\Phi_2^\dagger\Phi_1 \\
     i(\Phi_2^\dagger\Phi_1-\Phi_1^\dagger\Phi_2) \\
     \Phi_1^\dagger\Phi_1-\Phi_2^\dagger\Phi_2 \\
    \end{pmatrix}.
\end{equation}
By comparing the potential in the blinear notation (Eq.~\ref{eq:Vbilinear}) with the tranditional notation (Eq.~\ref{eq:V2hdm}), we can explicitly relate these two sets of parameters,
\begin{eqnarray}
\xi_{0} &\equiv& \frac{1}{2}(m_{11}^{2}+m_{22}^{2}), \qquad
\eta_{00} =(\lambda_{1}+\lambda_{2}+2 \lambda_{3})/8, \nonumber\\
\vec{\xi} &=& \left(- \Re\left(m_{12}^{2}\right), \Im\left(m_{12}^{2}\right), \tfrac{1}{2}(m_{11}^{2}-m_{22}^{2})\right)^{T},\nonumber\\
\vec{\eta} &=&\left(
\Re(\lambda_{6}+\lambda_{7})/4,-\Im(\lambda_{6}+\lambda_{7})/4, (\lambda_{1}-\lambda_{2})/8\right)^{T},\nonumber\\
E&=&\frac{1}{4}\left(\begin{array}{ccc}
\lambda_{4}+\Re\left(\lambda_{5}\right) & -\Im\left(\lambda_{5}\right) & \Re\left(\lambda_{6}-\lambda_{7}\right) \\
-\Im\left(\lambda_{5}\right) & \lambda_{4}-\Re\left(\lambda_{5}\right) & \Im\left(\lambda_{7}-\lambda_{6}\right) \\
\Re\left(\lambda_{6}-\lambda_{7}\right) & \Im\left(\lambda_{7}-\lambda_{6}\right) & \left(\lambda_{1}+\lambda_{2}-2 \lambda_{3}\right)/2
\end{array}\right).
\end{eqnarray}

In the 4-dimensional orbit space, the physical region is confined to the interior of the forward light-cone, i.e., $K_0 \geqslant |\vec{K}|$.
Because $K^\mu$ can be decomposed from $\underline{K}_{ij}=\Phi_i^\dagger \Phi_j$, by definition:
\begin{equation}
    \underline{K}\equiv
    \begin{pmatrix}
      \Phi_1^\dagger \Phi_1 & \Phi_2^\dagger \Phi_1\\
      \Phi_1^\dagger \Phi_2 & \Phi_2^\dagger \Phi_2\\
    \end{pmatrix}
    \equiv
    \frac{1}{2}\begin{pmatrix}
      K_0+K_3 & K_1-iK_2\\
      K_1+iK_2 & K_0-K_3\\
    \end{pmatrix},
\end{equation}
and the matrix $\underline{K}$ is actually a semi-positive matrix when $\Phi_i=(\phi_{i\uparrow},\phi_{i\downarrow})^T$ are $SU(2)_L$ doublets,
\begin{equation}\label{eq:Kphiphi}
    \underline{K}=\underline{\phi} \underline{\phi}^\dagger,\quad
    \underline{\phi}=\begin{pmatrix}
      \phi_{1\uparrow} & \phi_{1\downarrow}\\
      \phi_{2\downarrow} & \phi_{2\downarrow}
    \end{pmatrix},
\end{equation}
which directly leads to
\begin{equation}\label{eq:lightcone} %{{{
\begin{cases}
  ~~\text{tr} \underline{K}=K_0 \geqslant 0,\\
  \det \underline{K}=(K_0^2-|\vec K|^2)/4 \geqslant 0.
\end{cases}
\end{equation}%}}}
Therefore in the bilinear notation, the tree-level 2HDM scalar potential is a real quadratic function of $(K_0,\vec{K})$, and the physical region is defined inside the forward light-cone.

\subsection{$Z_2$ symmetry in the bilinear notation} \label{sub:softz2}
The $Z_2$ symmetry is imposed on the 2HDM by assigning $Z_2$ charges to scalar and fermion fields. In Eq.~\eqref{eq:V2hdm}, the two Higgs doublets $\Phi_{1}$ and $\Phi_2$ carry the $Z_2$ charges of $-1$ and $+1$ respectively, forbidding the $(\Phi_1^{\dagger }\Phi_1) (\Phi_1^{\dagger }\Phi_2)$ and $ (\Phi_2^{\dagger }\Phi_2) (\Phi_1^{\dagger }\Phi_2)$ terms in the potential.

As for the Yukawa interactions, fermions are also assigned with negative or positive $Z_2$ charges, then forced to interact with only $\Phi_1$ or $\Phi_2$. Usually, the patterns of $Z_2$ charges assignments are divided into four types~\cite{Haber:1978jt,Donoghue:1978cj,Hall:1981bc,Barger:1989fj,Barnett:1983mm,Barnett:1984zy}: Type I, Type II, Type X and Type Y, as listed in Table~\ref{tab:z2charge}. For fermions with different $Z_2$ charges, the vectors $\vec{Y}$'s projected by their Yukawa couplings are opposite to each other. For example, in the orbit space of $Z_2$ eigenbasis $(\Phi_1,\Phi_2)$, the Yukawa coupling of fermion with positive $Z_2$ charge yield $Y^\mu\propto(1,0,0,-1)$ and the Yukawa coupling of fermion with negative $Z_2$ charge yield $Y^\mu\propto(1,0,0,1)$.

\begin{table}
    \centering
    \caption{$Z_2$ charge assignment for different types}
    \begin{tabular}{|c|c|c|c|c|c|c|}
    \hline
         & $u_R$ & $d_R$ & $l_R$ & $\Phi_1$ & $\Phi_2$ & Directions of $\hat{Y}_f$ \\
    \hline
        Type I  & $+$ & $+$ & $+$ & $-$ & $+$ & $\hat{Y}_u=\hat{Y}_d=\hat{Y}_l$  \\
    \hline 
        Type II & $+$ & $-$ & $-$ & $-$ & $+$ & $-\hat{Y}_u=\hat{Y}_d=\hat{Y}_l$ \\
    \hline 
        Type X  & $+$ & $+$ & $-$ & $-$ & $+$ & $\hat{Y}_u=\hat{Y}_d=-\hat{Y}_l$ \\
    \hline 
        Type Y  & $+$ & $-$ & $+$ & $-$ & $+$ & $\hat{Y}_u=-\hat{Y}_d=\hat{Y}_l$ \\
    \hline
    \end{tabular}
    \label{tab:z2charge}
\end{table}

\subsection{Tensor notation}
For the completeness of this paper, here we reviewed another basis invariant notation to analyze the 2HDM potential, the tensor notation~\cite{JforScalarFermion,Branco:2005em,Gunion:2005ja}. It is straightforward to express the 2HDM scalar potential in an $U(2)_\Phi$ basis invariant form,
\begin{equation}\label{eq:Vtensor}
    V=\mu_{ij}\Phi_i^\dagger \Phi_j +\lambda_{ij,kl}(\Phi_i^\dagger \Phi_j)(\Phi_k^\dagger \Phi_l).
\end{equation}
As a result $\mu_{ij}$ and $\lambda_{ij,kl}$ transform covariantly with $\Phi_i$ under the $U(2)_\Phi$ basis transformation,
\begin{equation}\label{eq:tensorcoefficientTransform}
    \mu'_{ij}=U_{ik}\mu_{kl} U^*_{jl}, \quad
    \lambda'_{ij,kl}= U_{ip} U_{kr} \lambda_{pq,rs} U^*_{jq}U^*_{ls}.
\end{equation}
By definition, $\lambda_{ij,kl}=\lambda_{kl,ij}$, and hermiticity requires that $\mu_{ij}=\mu_{ji}^*,~\lambda_{kl,ij}=\lambda_{lk,ji}^*$. Under the basis of Eq.~\eqref{eq:V2hdm}, we have the following relations explicitly,
\begin{equation}
\begin{array}{rl}
\mu_{11}=m_{11}^{2}, & \quad \mu_{22}=m_{22}^{2}, \\
\mu_{12}=-m_{12}^{2}, & \quad \mu_{21}=-m_{12}^{2}{ }^{*} \\
\lambda_{11,11}=\lambda_{1}, & \quad \lambda_{22,22}=\lambda_{2}, \\
\lambda_{11,22}=\lambda_{22,11}=\lambda_{3}, & \quad \lambda_{12,21}=\lambda_{21,12}=\lambda_{4}, \\
\lambda_{12,12}=\lambda_{5}, & \quad \lambda_{21,21}=\lambda_{5}^{*}, \\
\lambda_{11,12}=\lambda_{12,11}=\lambda_{6}, & \quad \lambda_{11,21}=\lambda_{21,11}=\lambda_{6}^{*}, \\
\lambda_{22,12}=\lambda_{12,22}=\lambda_{7}, & \quad \lambda_{22,21}=\lambda_{21,22}=\lambda_{7}^{*}.
\end{array}
\end{equation}

The potential is invariant under the GCP symmetry~Eq.~\eqref{eq:GCP} when $\mu_{ij}$ and $\lambda_{ij,kl}$ satisfy
\begin{equation}
    \mu_{ij} = X_{ik}\mu_{kl}^* X_{lj}^*,
    \quad
    \lambda_{ij,kl}= X_{im} X_{kn} \lambda_{mp,nq}^* X_{jp}^* X_{lq}^*.
\end{equation}

One can construct several CP invariants to determine whether a potential is GCP invariant~\cite{JforScalarFermion}.
Similar to the Jarlskog invariant~\cite{Jarlskog:1985ht}, a $SU(3)_{L/R}$ invariant in quark family space, the CP invariants of 2HDM scalar potential are constructed from tensor products of $\mu_{ij}$ and $\lambda_{ij,kl}$ as $U(2)_\Phi$ invariants in scalar family space.
And tensor notation can also be used to construct CP invariants for scalar fermion interaction after extending tensor structures to fermion family space~\cite{JforScalarFermion}.
In addition, a recent development in tensor notation is using the Hilbert series to systematically construct all possible CP invariants~\cite{Trautner:2018ipq}, and similar procedures can also be used to construct CP invariant in the lepton sector with Majorana terms~\cite{Yu:2021cco}.

\section{Effective potential from Scalar Loop Contribution}
\label{appsub:scalar}
Here we show the calculation of the effective potential from scalar loop contribution in detail. We employ the notations in Ref.~\cite{Degee:2009vp} to link the eight scalar fields $\varphi_i$ with the bilinear forms $K^\mu$,
\begin{equation}
 \begin{aligned}
    & \mathcal{L}=\Omega_{\mu}\left(\partial_{\alpha} \Phi_i\right)^{\dagger} \sigma_{ij}^{\mu}\left(\partial^{\alpha} \Phi_j\right) -V,\quad \Omega^2=1, \\
    &V_{\rm tree}=\xi_{\mu} K^{\mu}+\eta_{\mu \nu} K^{\mu} K^{\nu},\\
    &\varphi_{a}=\left(\operatorname{Re} \phi_{1, \uparrow}, \operatorname{Im} \phi_{1, \uparrow}, \operatorname{Re} \phi_{2, \uparrow}, \operatorname{Im} \phi_{2, \uparrow}, \operatorname{Re} \phi_{1, \downarrow}, \operatorname{Im} \phi_{1, \downarrow}, \operatorname{Re} \phi_{2, \downarrow}, \operatorname{Im} \phi_{2, \downarrow}\right),\\
    &K^{\mu}=\varphi_{a} \Sigma_{a b}^{\mu} \varphi_{b},\\
    &(\Omega_\rho\Sigma^\rho)^{-1}=\Omega_{\rho}\bar{\Sigma}^{\rho}.
 \end{aligned}
\end{equation}
Note that $\Omega_\mu=(1,0,0,0)$ for the canonical kinetic term 
The matrix $\bar{\Sigma}^\mu=(\Sigma^0,-\Sigma^i)$ and the $8\times8$ symmetric matrics $\Sigma^\mu$ defined in Eq.~\eqref{sigma} are  
\begin{equation}
    \Sigma^\mu=\Sigma^\mu_4\oplus\Sigma^\mu_4,\quad
    \Sigma_4^0=\mathbb{1}_4,\quad
    \Sigma_4^1=\begin{pmatrix}
     0 & \mathbb{1}_2\\
     \mathbb{1}_2 & 0\\
    \end{pmatrix},\quad
    \Sigma_4^2=\begin{pmatrix}
     0 & \mathbb{i}_2\\
     -\mathbb{i}_2 & 0\\
    \end{pmatrix},\quad
    \Sigma_4^3=\begin{pmatrix}
     \mathbb{1}_2 & 0\\
     0 & -\mathbb{1}_2\\
    \end{pmatrix},
\end{equation}
where $\mathbb{1}_d$ is the $d\times d$ identity matrix and $\mathbb{i}_2\equiv\left(\begin{smallmatrix} 0&1\\-1&0 \end{smallmatrix}\right)$. Because $(\mathbb{i}_2)^2 = -\mathbb{1}_2$, the matrix $\Sigma^\mu$ share the same algebra with the pauli matrix $\sigma^\mu$, e.g.,
\begin{align}
\label{eq:Sigmaalgebra}
      [\Sigma^i ,\Sigma^j] = 2\mathbb{i}_8 \epsilon^{ijk} \Sigma^k &, \quad (\vec{w}\cdot\vec\Sigma)^2=|\vec{w}|^2\mathbb{1}_{8}, \\
      \frac{1}{2}(\bar{\Sigma}^{\mu}\Sigma^{\nu}+\bar{\Sigma}^{\nu}\Sigma^{\mu})&=g^{\mu\nu}\mathbb{1}_{8},\\
      \Sigma^{\mu}\bar{\Sigma}^{\rho}\Sigma^{\nu}=g^{\mu\rho}\Sigma^{\nu}+g^{\rho\nu}\Sigma^{\mu}&-g^{\mu\nu}\Sigma^{\rho}+\mathbb{i}_8 \epsilon^{\mu\rho\nu}_{\quad\lambda}\Sigma^{\lambda},\\
      \bar{\Sigma}^{\mu}\Sigma^{\rho}\bar{\Sigma}^{\nu}=g^{\mu\rho}\bar{\Sigma}^{\nu}+g^{\rho\nu}\bar{\Sigma}^{\mu}&-g^{\mu\nu}\bar{\Sigma}^{\rho}-\mathbb{i}_8\epsilon^{\mu\rho\nu}_{\quad\lambda}\bar{\Sigma}^{\lambda}.
\end{align}
Here $\mathbb{i}_8 \equiv \mathbb{1}_4\otimes\mathbb{i}_2$ is an anti-symmetric matrix who commutes with $\Sigma^\mu$ and satisfies $(\mathbb{i}_8)^2=-\mathbb{1}_8$, and $\vec{w}$ is an arbitrary vector. These identities help to translate some expressions of $\varphi_a$ to bilinear forms. For example,
\begin{align}
    \varphi\mathbb{i}_8\Sigma^\mu\varphi =0,    \quad \varphi\Sigma^{\mu}\bar{\Sigma}^{\rho}\Sigma^{\nu}\varphi =g^{\mu\rho}K^{\nu}+g^{\rho\nu}K^{\mu}-g^{\mu\nu}K^{\rho}.
\end{align}
Then we evaluate the second derivative of $\mathcal{L}$
\begin{equation}
    -\frac{\delta^{2}\mathcal{L}}{\delta\varphi_{a}\delta\varphi_{b}}=\Omega_{\rho}\Sigma_{ab}^{\rho}\partial^{2}+\xi_{\mu}\Sigma_{ab}^{\mu}+2\eta_{\mu\nu}(\varphi_{c}\Sigma_{cd}^{\mu}\varphi_{d})\Sigma_{ab}^{\nu}+
    4\eta_{\mu\nu}\Sigma^{\mu}_{ac}(\varphi_c\varphi_d)\Sigma^{\nu}_{db}.
\end{equation}
In the following, we work in the frame with the canonical kinetic term with $\Omega_\mu=(1,0,0,0)$, and the scalar mass matrix is
\begin{equation}
 \begin{aligned}
        &\mathbf{M}_S^2(\varphi)_{ab}=A_{ab}+B_{ab},\\
        &A_{ab}=A_\mu\Sigma_{ab}^\mu,
        ~\quad A_{\mu}=2\eta_{\mu\nu}K^{\nu}+\xi_{\mu},\\
        &B_{ab}=4\eta_{\mu\nu}\Sigma_{ac}^\mu\varphi_c\varphi_d\Sigma_{db}^\nu.
  \label{eq:log}
 \end{aligned}
\end{equation}
To deal with $\mathbf{Tr}(\mathbf{M}_S^{2n})$ in Eq.~\eqref{eq:Vcws}, we expand the binomial 
\begin{equation}\label{eq:expandABn}
    \mathbf{Tr}[(A_{ab}+B_{ab})^{n}]=\sum_{l=0}^{n}\sum_{\{p_i\}}^{\sum p_i=n-l}N_s(\{p_i\})\mathbf{Tr}(A^{p_{1}}BA^{p_{2}}B\cdots A^{p_{l}}B).
\end{equation}
And we need to evaluate $(A_\mu\Sigma^\mu)^p$. Using the identities in Eq.~\eqref{eq:Sigmaalgebra},
\begin{align}
(A_\mu\Sigma^\mu)^p &= (A_0\mathbb{1}_8+\vec{A}\cdot\vec{\Sigma})^p,   \nn\\
    &= \sum_{k=0}^p C_p^k (A_0)^{p-k} (\vec{A}\cdot\vec{\Sigma})^k,   \nn\\
    &= \sum_{k=0}^{p/2} C_p^{2k} (A_0)^{p-2k} |\vec{A}|^{2k} \mathbb{1}_8
     + \sum_{k=0}^{(p-1)/2} C_p^{2k+1} (A_0)^{p-2k-1}|\vec{A}|^{2k} (\vec{A}\cdot\vec{\Sigma}),
\end{align}
where $C_p^k$ is the binomial coefficient and
\begin{align}
A_0&=2\eta_{00}K_0+2\vec{\eta}\cdot\vec{K} + \xi_0,\\
\vec{A}&=2K_0\vec{\eta}+2E\vec{K} + \vec\xi.
\end{align}
For simplicity, we define a new four-vector $F(p)_\mu$ from $(A_0,\vec{A})$
\begin{align}
 F(p)_0 &\equiv
 \sum_{k=0}^{p/2} C_p^{2k} (A_0)^{p-2k} |\vec{A}|^{2k},\\
 \vec{F}(p) &\equiv\left\{
 \begin{aligned}
 - &\sum_{k=0}^{(p-1)/2} C_p^{2k+1} (A_0)^{p-2k-1}|\vec{A}|^{2k} \vec{A}\qquad &(p\neq0),\\
 &0\qquad &(p=0),
 \end{aligned}
 \right.
\end{align}
and we have
\begin{equation}
    (A_\mu\Sigma^\mu)^p = F(p)_\mu \bar{\Sigma}^\mu.
\end{equation}
The series in Eq.~\eqref{eq:expandABn} are then calculated as
\[ 
\begin{split}
\mathbf{Tr}(A^{p_{1}}BA^{p_{2}}B\cdots A^{p_{l}}B)&
=4^l\eta_{\mu_1\nu_1}\cdots\eta_{\mu_l\nu_l} \prod_{i=1}^l A(p_i)_{\rho_i} \varphi \Sigma^{\nu_i}\bar{\Sigma}^{\rho_i}\Sigma^{\mu_{i+1}} \varphi\\
&=4^l\eta_{\mu_1\nu_1}\cdots\eta_{\mu_l\nu_l} \prod_{i=1}^l  S_{p_i}^{\nu_i\mu_{i+1}}\\
&=4^l \mathbf{tr}\left(\eta \cdot S_{p_1} \cdots \eta \cdot S_{p_l}\right),\\
\end{split}
\]
where $\mu_{l+1}\equiv\mu_1$ and the trace $\mathbf{tr}$ is taken in the orbit space. The symmetric tensor $S_p^{\mu\nu}=F(p)_\rho \varphi\Sigma^\mu \bar{\Sigma}^\rho \Sigma^\nu \varphi =F(p)^{\mu}K^{\nu}+F(p)^{\nu}K^{\mu}-g^{\mu\nu}(F(p)K) $. And the effective potential can be expressed as
\footnote{For simplicity, the $\ln p_E$ is dropped here.}
\begin{align}
    V_{\rm CW}^{(S)} &=\frac{1}{2}\int
\frac{d^4p_E}{2\pi^4}\left[\mathbf{Tr}\sum_{n=1}^\infty\frac{1}{n}\left(-\frac{\mathbf{M}_S^2}{p^2_E}\right)^n\right] \nonumber\\
    &=\frac{1}{2}\int \frac{d^4p_E}{2\pi^4} \sum_n (-)^n\frac{1}{n(p_E^2)^n}\sum_{l=0}^{n}\sum_{\{p_i\}}^{\sum p_i=n-l}N_s(\{p_i\})\mathbf{Tr}(A^{p_{1}}BA^{p_{2}}B\cdots A^{p_{l}}B) \nonumber\\
    &=\frac{1}{2}\int \frac{d^4p_E}{2\pi^4} \sum_n (-)^n\frac{1}{n(p_E^2)^n}\sum_{l=0}^{n}\sum_{\{p_i\}}^{\sum p_i=n-l}N_s(\{p_i\}) \mathbf{tr}\left(\eta \cdot S_{p_1} \cdots \eta \cdot S_{p_l}\right)
\end{align}

In the end, the $V_{\rm CW}^{(S)}$ is expressed as a series defined in the orbit space.
It is worth mentioning that the discussion of CP property is independent of regularization.
When the potential is CP-even, as we discussed, we can apply the CP transformation before and after the regularization and nothing will change. Finally, we can conclude that the CP property of the (CP conserving) potential tree-level potential is not violated by the Coleman-Weinberg potential from scalar loop contribution.

\section{Renormalization Conditions}\label{app:renormalization}
To compare with the renormalization conditions in Ref.~\cite{Basler:2018cwe}, we follow their notations and the field expanded around the vacuum $v_1,v_2$ are
\begin{equation}
\Phi_1 =\frac{1}{\sqrt{2}}\left(\begin{array}{c}
\rho_1+\mathrm{i} \eta_1 \\
v_1+\zeta_1+\mathrm{i} \psi_1
\end{array}\right),\quad
\Phi_2 =\frac{1}{\sqrt{2}}\left(\begin{array}{c}
\rho_2+\mathrm{i} \eta_2 \\
v_2+\zeta_2+\mathrm{i}\psi_2
\end{array}\right).
\end{equation}
The renormalization conditions are
\begin{align}
\label{eq:renormalizationphii}
    &\partial_{\varphi_a}(V_{\rm CT}+V_{\rm CW})\big|_{\varphi_a=\langle\varphi_a\rangle_{\rm tree}}=0,\\
\label{eq:renormalizationphii2}
    &\partial_{\varphi_a}\partial_{\varphi_b}(V_{\rm CT}+V_{\rm CW})\big|_{\varphi_a=\langle\varphi_a\rangle_{\rm tree}}=0.\\
   &\varphi_a \equiv\left\{\rho_1, \eta_1, \zeta_1, \psi_1, \rho_2, \eta_2, \zeta_2, \psi_2\right\},~
   \langle\varphi_a\rangle_{\rm tree} =\left\{0,0,v_1,0, 0,0,v_2,0\right\}.\nonumber
\end{align}
Naively, there are 8+36 renormalization conditions from Eqs.~\eqref{eq:renormalizationphii} and~\eqref{eq:renormalizationphii2}. However, for any function of the form $f(\Phi^\dagger_i\Phi_j)$, its first and second derivative
satisfy some identities so that most of the renormalization conditions are redundant.

We have the following 5 identities for the first derivatives,
\begin{align}
    & \partial_{\rho_1}=0, \\
    & \partial_{\rho_2}=0, \\
    & \partial_{\eta_1}=0, \\
    & \partial_{\eta_2}=0, \\
    & c_\beta \partial_{\psi_1} + s_\beta \partial_{\psi_2}=0,
\end{align}
where $\partial_{\phi_i}=0$ denotes $\partial_{\phi_i}f|_{\phi=\langle\phi\rangle_{\rm tree}}=0$ for any function $f(\Phi^\dagger_i\Phi_j)$ and $\tan\beta=v_2/v_1$.
Therefore, we are left with 3 independent renormalization conditions from Eq.~\eqref{eq:renormalizationphii},
\begin{align}
    \label{eq:dVVphi1}
    & \partial_{\zeta_1}\VV=0, \\
    & \partial_{\zeta_2}\VV=0, \\
    \label{eq:dVVphi3}
    & \dphi{8}\VV=0.
\end{align}

We have the following 26 identities for the second derivatives,
\begin{align}
    & \dphi{1}\dphi{2}=0, \\
    & \dphi{1}\dphi{3}=0, \\
    & \dphi{1}\dphi{4}=0, \\
    & \dphi{1}\dphi{7}=0, \\
    & \dphi{1}\dphi{8}=0, \\
    & \dphi{2}\dphi{3}=0, \\
    & \dphi{2}\dphi{4}=0, \\
    & \dphi{2}\dphi{7}=0, \\
    & \dphi{2}\dphi{8}=0, \\
    & \dphi{3}\dphi{4}=0, \\
    & \dphi{3}\dphi{5}=0, \\
    & \dphi{3}\dphi{6}=0, \\
    & \dphi{4}\dphi{5}=0, \\
    & \dphi{4}\dphi{6}=0, \\
    & \dphi{5}\dphi{6}=0, \\
    & \dphi{5}\dphi{7}=0, \\
    & \dphi{5}\dphi{8}=0, \\
    & \dphi{6}\dphi{7}=0, \\
    & \dphi{6}\dphi{8}=0, \\
    & \dphi{1}\dphi{1}=\dphi{2}\dphi{2}, \\
    & \dphi{1}\dphi{1}=\dphi{4}\dphi{4}, \\
    & \dphi{5}\dphi{5}=\dphi{6}\dphi{6}, \\
    & \dphi{1}\dphi{5}=\dphi{2}\dphi{6}, \\
    & \dphi{1}\dphi{5}=\dphi{4}\dphi{8}, \\
    & \dphi{2}\dphi{5}=-\dphi{1}\dphi{6}, \\
    & \dphi{2}\dphi{5}=\dphi{4}\dphi{7}.
\end{align}
Then, there are 10 independent renormalization conditions from the second derivatives. However, three of them are satisfied automatically when the renormalization conditions from the first derivatives are satisfied, because of the following identities,
\begin{align}
    & \dphi{1}^2=\frac{1}{2v}\dphi{3},\\
    & \dphi{1}\dphi{5}= \frac{1}{2v}\dphi{7},\\
    & \dphi{1}\dphi{6}= \frac{1}{2v}\dphi{8}.
\end{align}
Finally, we are left with only 7 independent renormalization conditions from Eq.~\eqref{eq:renormalizationphii2},
\begin{align}
\label{eq:ddVVphi1}
    & \dphi{5}^2\VV=0,\\
    & \dphi{3}^2\VV=0,\\
    & \dphi{7}^2\VV=0,\\
    & \dphi{8}^2\VV=0,\\
    & \dphi{3}\dphi{7}\VV=0,\\
    & \dphi{3}\dphi{8}\VV=0,\\
\label{eq:ddVVphi7}
    & \dphi{7}\dphi{8}\VV=0.
\end{align}
And we have 10 independent renormalization conditions from Eqs.~\eqref{eq:dVVphi1}-\eqref{eq:dVVphi3} and Eqs.~\eqref{eq:ddVVphi1}-\eqref{eq:ddVVphi7} in total.

\bibliographystyle{apsrev4-1}
\bibliography{ref}

\providecommand{\noopsort}[1]{}\providecommand{\singleletter}[1]{#1}%
\begin{thebibliography}{52}%
\makeatletter
\providecommand \@ifxundefined [1]{%
 \@ifx{#1\undefined}
}%
\providecommand \@ifnum [1]{%
 \ifnum #1\expandafter \@firstoftwo
 \else \expandafter \@secondoftwo
 \fi
}%
\providecommand \@ifx [1]{%
 \ifx #1\expandafter \@firstoftwo
 \else \expandafter \@secondoftwo
 \fi
}%
\providecommand \natexlab [1]{#1}%
\providecommand \enquote  [1]{``#1''}%
\providecommand \bibnamefont  [1]{#1}%
\providecommand \bibfnamefont [1]{#1}%
\providecommand \citenamefont [1]{#1}%
\providecommand \href@noop [0]{\@secondoftwo}%
\providecommand \href [0]{\begingroup \@sanitize@url \@href}%
\providecommand \@href[1]{\@@startlink{#1}\@@href}%
\providecommand \@@href[1]{\endgroup#1\@@endlink}%
\providecommand \@sanitize@url [0]{\catcode `\\12\catcode `\$12\catcode
  `\&12\catcode `\#12\catcode `\^12\catcode `\_12\catcode `\%12\relax}%
\providecommand \@@startlink[1]{}%
\providecommand \@@endlink[0]{}%
\providecommand \url  [0]{\begingroup\@sanitize@url \@url }%
\providecommand \@url [1]{\endgroup\@href {#1}{\urlprefix }}%
\providecommand \urlprefix  [0]{URL }%
\providecommand \Eprint [0]{\href }%
\providecommand \doibase [0]{http://dx.doi.org/}%
\providecommand \selectlanguage [0]{\@gobble}%
\providecommand \bibinfo  [0]{\@secondoftwo}%
\providecommand \bibfield  [0]{\@secondoftwo}%
\providecommand \translation [1]{[#1]}%
\providecommand \BibitemOpen [0]{}%
\providecommand \bibitemStop [0]{}%
\providecommand \bibitemNoStop [0]{.\EOS\space}%
\providecommand \EOS [0]{\spacefactor3000\relax}%
\providecommand \BibitemShut  [1]{\csname bibitem#1\endcsname}%
\let\auto@bib@innerbib\@empty
%</preamble>
\bibitem [{\citenamefont {Lee}(1973)}]{Lee:1973iz}%
  \BibitemOpen
  \bibfield  {author} {\bibinfo {author} {\bibfnamefont {T.~D.}\ \bibnamefont
  {Lee}},\ }\href {\doibase 10.1103/PhysRevD.8.1226} {\bibfield  {journal}
  {\bibinfo  {journal} {Phys. Rev. D}\ }\textbf {\bibinfo {volume} {8}},\
  \bibinfo {pages} {1226} (\bibinfo {year} {1973})}\BibitemShut {NoStop}%
\bibitem [{\citenamefont {Botella}\ and\ \citenamefont
  {Silva}(1995)}]{JforScalarFermion}%
  \BibitemOpen
  \bibfield  {author} {\bibinfo {author} {\bibfnamefont {F.~J.}\ \bibnamefont
  {Botella}}\ and\ \bibinfo {author} {\bibfnamefont {J.~P.}\ \bibnamefont
  {Silva}},\ }\href {\doibase 10.1103/PhysRevD.51.3870} {\bibfield  {journal}
  {\bibinfo  {journal} {Phys. Rev. D}\ }\textbf {\bibinfo {volume} {51}},\
  \bibinfo {pages} {3870} (\bibinfo {year} {1995})},\ \Eprint
  {http://arxiv.org/abs/hep-ph/9411288} {arXiv:hep-ph/9411288} \BibitemShut
  {NoStop}%
\bibitem [{\citenamefont {Branco}\ \emph {et~al.}(2005)\citenamefont {Branco},
  \citenamefont {Rebelo},\ and\ \citenamefont {Silva-Marcos}}]{Branco:2005em}%
  \BibitemOpen
  \bibfield  {author} {\bibinfo {author} {\bibfnamefont {G.~C.}\ \bibnamefont
  {Branco}}, \bibinfo {author} {\bibfnamefont {M.~N.}\ \bibnamefont {Rebelo}},
  \ and\ \bibinfo {author} {\bibfnamefont {J.~I.}\ \bibnamefont
  {Silva-Marcos}},\ }\href {\doibase 10.1016/j.physletb.2005.03.075} {\bibfield
   {journal} {\bibinfo  {journal} {Phys. Lett. B}\ }\textbf {\bibinfo {volume}
  {614}},\ \bibinfo {pages} {187} (\bibinfo {year} {2005})},\ \Eprint
  {http://arxiv.org/abs/hep-ph/0502118} {arXiv:hep-ph/0502118} \BibitemShut
  {NoStop}%
\bibitem [{\citenamefont {Gunion}\ and\ \citenamefont
  {Haber}(2005)}]{Gunion:2005ja}%
  \BibitemOpen
  \bibfield  {author} {\bibinfo {author} {\bibfnamefont {J.~F.}\ \bibnamefont
  {Gunion}}\ and\ \bibinfo {author} {\bibfnamefont {H.~E.}\ \bibnamefont
  {Haber}},\ }\href {\doibase 10.1103/PhysRevD.72.095002} {\bibfield  {journal}
  {\bibinfo  {journal} {Phys. Rev. D}\ }\textbf {\bibinfo {volume} {72}},\
  \bibinfo {pages} {095002} (\bibinfo {year} {2005})},\ \Eprint
  {http://arxiv.org/abs/hep-ph/0506227} {arXiv:hep-ph/0506227} \BibitemShut
  {NoStop}%
\bibitem [{\citenamefont {Trautner}(2019)}]{Trautner:2018ipq}%
  \BibitemOpen
  \bibfield  {author} {\bibinfo {author} {\bibfnamefont {A.}~\bibnamefont
  {Trautner}},\ }\href {\doibase 10.1007/JHEP05(2019)208} {\bibfield  {journal}
  {\bibinfo  {journal} {JHEP}\ }\textbf {\bibinfo {volume} {05}},\ \bibinfo
  {pages} {208} (\bibinfo {year} {2019})},\ \Eprint
  {http://arxiv.org/abs/1812.02614} {arXiv:1812.02614 [hep-ph]} \BibitemShut
  {NoStop}%
\bibitem [{\citenamefont {Cline}\ \emph {et~al.}(2011)\citenamefont {Cline},
  \citenamefont {Kainulainen},\ and\ \citenamefont {Trott}}]{Cline:2011mm}%
  \BibitemOpen
  \bibfield  {author} {\bibinfo {author} {\bibfnamefont {J.~M.}\ \bibnamefont
  {Cline}}, \bibinfo {author} {\bibfnamefont {K.}~\bibnamefont {Kainulainen}},
  \ and\ \bibinfo {author} {\bibfnamefont {M.}~\bibnamefont {Trott}},\ }\href
  {\doibase 10.1007/JHEP11(2011)089} {\bibfield  {journal} {\bibinfo  {journal}
  {JHEP}\ }\textbf {\bibinfo {volume} {11}},\ \bibinfo {pages} {089} (\bibinfo
  {year} {2011})},\ \Eprint {http://arxiv.org/abs/1107.3559} {arXiv:1107.3559
  [hep-ph]} \BibitemShut {NoStop}%
\bibitem [{\citenamefont {Basler}\ \emph {et~al.}(2017)\citenamefont {Basler},
  \citenamefont {Krause}, \citenamefont {Muhlleitner}, \citenamefont
  {Wittbrodt},\ and\ \citenamefont {Wlotzka}}]{Basler:2016obg}%
  \BibitemOpen
  \bibfield  {author} {\bibinfo {author} {\bibfnamefont {P.}~\bibnamefont
  {Basler}}, \bibinfo {author} {\bibfnamefont {M.}~\bibnamefont {Krause}},
  \bibinfo {author} {\bibfnamefont {M.}~\bibnamefont {Muhlleitner}}, \bibinfo
  {author} {\bibfnamefont {J.}~\bibnamefont {Wittbrodt}}, \ and\ \bibinfo
  {author} {\bibfnamefont {A.}~\bibnamefont {Wlotzka}},\ }\href {\doibase
  10.1007/JHEP02(2017)121} {\bibfield  {journal} {\bibinfo  {journal} {JHEP}\
  }\textbf {\bibinfo {volume} {02}},\ \bibinfo {pages} {121} (\bibinfo {year}
  {2017})},\ \Eprint {http://arxiv.org/abs/1612.04086} {arXiv:1612.04086
  [hep-ph]} \BibitemShut {NoStop}%
\bibitem [{\citenamefont {Basler}\ \emph {et~al.}(2020)\citenamefont {Basler},
  \citenamefont {M\"uhlleitner},\ and\ \citenamefont
  {M\"uller}}]{Basler:2019iuu}%
  \BibitemOpen
  \bibfield  {author} {\bibinfo {author} {\bibfnamefont {P.}~\bibnamefont
  {Basler}}, \bibinfo {author} {\bibfnamefont {M.}~\bibnamefont
  {M\"uhlleitner}}, \ and\ \bibinfo {author} {\bibfnamefont {J.}~\bibnamefont
  {M\"uller}},\ }\href {\doibase 10.1007/JHEP05(2020)016} {\bibfield  {journal}
  {\bibinfo  {journal} {JHEP}\ }\textbf {\bibinfo {volume} {05}},\ \bibinfo
  {pages} {016} (\bibinfo {year} {2020})},\ \Eprint
  {http://arxiv.org/abs/1912.10477} {arXiv:1912.10477 [hep-ph]} \BibitemShut
  {NoStop}%
\bibitem [{\citenamefont {Ferreira}\ \emph {et~al.}(2020)\citenamefont
  {Ferreira}, \citenamefont {Morrison},\ and\ \citenamefont
  {Profumo}}]{Ferreira:2019bij}%
  \BibitemOpen
  \bibfield  {author} {\bibinfo {author} {\bibfnamefont {P.~M.}\ \bibnamefont
  {Ferreira}}, \bibinfo {author} {\bibfnamefont {L.~A.}\ \bibnamefont
  {Morrison}}, \ and\ \bibinfo {author} {\bibfnamefont {S.}~\bibnamefont
  {Profumo}},\ }\href {\doibase 10.1007/JHEP04(2020)125} {\bibfield  {journal}
  {\bibinfo  {journal} {JHEP}\ }\textbf {\bibinfo {volume} {04}},\ \bibinfo
  {pages} {125} (\bibinfo {year} {2020})},\ \Eprint
  {http://arxiv.org/abs/1910.08662} {arXiv:1910.08662 [hep-ph]} \BibitemShut
  {NoStop}%
\bibitem [{\citenamefont {Branco}\ \emph {et~al.}(2012)\citenamefont {Branco},
  \citenamefont {Ferreira}, \citenamefont {Lavoura}, \citenamefont {Rebelo},
  \citenamefont {Sher},\ and\ \citenamefont {Silva}}]{Branco:2011iw}%
  \BibitemOpen
  \bibfield  {author} {\bibinfo {author} {\bibfnamefont {G.~C.}\ \bibnamefont
  {Branco}}, \bibinfo {author} {\bibfnamefont {P.~M.}\ \bibnamefont
  {Ferreira}}, \bibinfo {author} {\bibfnamefont {L.}~\bibnamefont {Lavoura}},
  \bibinfo {author} {\bibfnamefont {M.~N.}\ \bibnamefont {Rebelo}}, \bibinfo
  {author} {\bibfnamefont {M.}~\bibnamefont {Sher}}, \ and\ \bibinfo {author}
  {\bibfnamefont {J.~P.}\ \bibnamefont {Silva}},\ }\href {\doibase
  10.1016/j.physrep.2012.02.002} {\bibfield  {journal} {\bibinfo  {journal}
  {Phys. Rept.}\ }\textbf {\bibinfo {volume} {516}},\ \bibinfo {pages} {1}
  (\bibinfo {year} {2012})},\ \Eprint {http://arxiv.org/abs/1106.0034}
  {arXiv:1106.0034 [hep-ph]} \BibitemShut {NoStop}%
\bibitem [{\citenamefont {Maniatis}\ \emph
  {et~al.}(2008{\natexlab{a}})\citenamefont {Maniatis}, \citenamefont {von
  Manteuffel},\ and\ \citenamefont {Nachtmann}}]{THDMbilinear_CPV}%
  \BibitemOpen
  \bibfield  {author} {\bibinfo {author} {\bibfnamefont {M.}~\bibnamefont
  {Maniatis}}, \bibinfo {author} {\bibfnamefont {A.}~\bibnamefont {von
  Manteuffel}}, \ and\ \bibinfo {author} {\bibfnamefont {O.}~\bibnamefont
  {Nachtmann}},\ }\href {\doibase 10.1140/epjc/s10052-008-0712-5} {\bibfield
  {journal} {\bibinfo  {journal} {Eur. Phys. J. C}\ }\textbf {\bibinfo {volume}
  {57}},\ \bibinfo {pages} {719} (\bibinfo {year} {2008}{\natexlab{a}})},\
  \Eprint {http://arxiv.org/abs/0707.3344} {arXiv:0707.3344 [hep-ph]}
  \BibitemShut {NoStop}%
\bibitem [{\citenamefont {Maniatis}\ \emph {et~al.}(2006)\citenamefont
  {Maniatis}, \citenamefont {von Manteuffel}, \citenamefont {Nachtmann},\ and\
  \citenamefont {Nagel}}]{THDMbilinear}%
  \BibitemOpen
  \bibfield  {author} {\bibinfo {author} {\bibfnamefont {M.}~\bibnamefont
  {Maniatis}}, \bibinfo {author} {\bibfnamefont {A.}~\bibnamefont {von
  Manteuffel}}, \bibinfo {author} {\bibfnamefont {O.}~\bibnamefont
  {Nachtmann}}, \ and\ \bibinfo {author} {\bibfnamefont {F.}~\bibnamefont
  {Nagel}},\ }\href {\doibase 10.1140/epjc/s10052-006-0016-6} {\bibfield
  {journal} {\bibinfo  {journal} {Eur. Phys. J. C}\ }\textbf {\bibinfo {volume}
  {48}},\ \bibinfo {pages} {805} (\bibinfo {year} {2006})},\ \Eprint
  {http://arxiv.org/abs/hep-ph/0605184} {arXiv:hep-ph/0605184} \BibitemShut
  {NoStop}%
\bibitem [{\citenamefont {Ivanov}(2007)}]{Ivanov:2006yq}%
  \BibitemOpen
  \bibfield  {author} {\bibinfo {author} {\bibfnamefont {I.~P.}\ \bibnamefont
  {Ivanov}},\ }\href {\doibase 10.1103/PhysRevD.75.035001} {\bibfield
  {journal} {\bibinfo  {journal} {Phys. Rev. D}\ }\textbf {\bibinfo {volume}
  {75}},\ \bibinfo {pages} {035001} (\bibinfo {year} {2007})},\ \bibinfo {note}
  {[Erratum: Phys.Rev.D 76, 039902 (2007)]},\ \Eprint
  {http://arxiv.org/abs/hep-ph/0609018} {arXiv:hep-ph/0609018} \BibitemShut
  {NoStop}%
\bibitem [{\citenamefont {Ivanov}(2008)}]{Ivanov:2007de}%
  \BibitemOpen
  \bibfield  {author} {\bibinfo {author} {\bibfnamefont {I.~P.}\ \bibnamefont
  {Ivanov}},\ }\href {\doibase 10.1103/PhysRevD.77.015017} {\bibfield
  {journal} {\bibinfo  {journal} {Phys. Rev. D}\ }\textbf {\bibinfo {volume}
  {77}},\ \bibinfo {pages} {015017} (\bibinfo {year} {2008})},\ \Eprint
  {http://arxiv.org/abs/0710.3490} {arXiv:0710.3490 [hep-ph]} \BibitemShut
  {NoStop}%
\bibitem [{\citenamefont {Nishi}(2006)}]{Nishi:2006tg}%
  \BibitemOpen
  \bibfield  {author} {\bibinfo {author} {\bibfnamefont {C.~C.}\ \bibnamefont
  {Nishi}},\ }\href {\doibase 10.1103/PhysRevD.76.119901} {\bibfield  {journal}
  {\bibinfo  {journal} {Phys. Rev. D}\ }\textbf {\bibinfo {volume} {74}},\
  \bibinfo {pages} {036003} (\bibinfo {year} {2006})},\ \bibinfo {note}
  {[Erratum: Phys.Rev.D 76, 119901 (2007)]},\ \Eprint
  {http://arxiv.org/abs/hep-ph/0605153} {arXiv:hep-ph/0605153} \BibitemShut
  {NoStop}%
\bibitem [{\citenamefont {Nagel}(2004)}]{Nagel:2004sw}%
  \BibitemOpen
  \bibfield  {author} {\bibinfo {author} {\bibfnamefont {F.}~\bibnamefont
  {Nagel}},\ }\emph {\bibinfo {title} {{New aspects of gauge-boson couplings
  and the Higgs sector}}},\ \href@noop {} {Ph.D. thesis},\ \bibinfo  {school}
  {Heidelberg U.} (\bibinfo {year} {2004})\BibitemShut {NoStop}%
\bibitem [{\citenamefont {Ivanov}(2006)}]{Ivanov:2005hg}%
  \BibitemOpen
  \bibfield  {author} {\bibinfo {author} {\bibfnamefont {I.~P.}\ \bibnamefont
  {Ivanov}},\ }\href {\doibase 10.1016/j.physletb.2005.10.015} {\bibfield
  {journal} {\bibinfo  {journal} {Phys. Lett. B}\ }\textbf {\bibinfo {volume}
  {632}},\ \bibinfo {pages} {360} (\bibinfo {year} {2006})},\ \Eprint
  {http://arxiv.org/abs/hep-ph/0507132} {arXiv:hep-ph/0507132} \BibitemShut
  {NoStop}%
\bibitem [{\citenamefont {Cao}\ \emph {et~al.}(2023)\citenamefont {Cao},
  \citenamefont {Cheng},\ and\ \citenamefont {Xu}}]{Cao:2022rgh}%
  \BibitemOpen
  \bibfield  {author} {\bibinfo {author} {\bibfnamefont {Q.-H.}\ \bibnamefont
  {Cao}}, \bibinfo {author} {\bibfnamefont {K.}~\bibnamefont {Cheng}}, \ and\
  \bibinfo {author} {\bibfnamefont {C.}~\bibnamefont {Xu}},\ }\href {\doibase
  10.1103/PhysRevD.107.015016} {\bibfield  {journal} {\bibinfo  {journal}
  {Phys. Rev. D}\ }\textbf {\bibinfo {volume} {107}},\ \bibinfo {pages}
  {015016} (\bibinfo {year} {2023})},\ \Eprint
  {http://arxiv.org/abs/2201.02989} {arXiv:2201.02989 [hep-ph]} \BibitemShut
  {NoStop}%
\bibitem [{\citenamefont {Sartore}\ \emph {et~al.}(2022)\citenamefont
  {Sartore}, \citenamefont {Maniatis}, \citenamefont {Schienbein},\ and\
  \citenamefont {Herrmann}}]{Sartore:2022sxh}%
  \BibitemOpen
  \bibfield  {author} {\bibinfo {author} {\bibfnamefont {L.}~\bibnamefont
  {Sartore}}, \bibinfo {author} {\bibfnamefont {M.}~\bibnamefont {Maniatis}},
  \bibinfo {author} {\bibfnamefont {I.}~\bibnamefont {Schienbein}}, \ and\
  \bibinfo {author} {\bibfnamefont {B.}~\bibnamefont {Herrmann}},\ }\href
  {\doibase 10.1007/JHEP12(2022)051} {\bibfield  {journal} {\bibinfo  {journal}
  {JHEP}\ }\textbf {\bibinfo {volume} {12}},\ \bibinfo {pages} {051} (\bibinfo
  {year} {2022})},\ \Eprint {http://arxiv.org/abs/2208.13719} {arXiv:2208.13719
  [hep-ph]} \BibitemShut {NoStop}%
\bibitem [{\citenamefont {Basler}\ and\ \citenamefont
  {M\"uhlleitner}(2019)}]{Basler:2018cwe}%
  \BibitemOpen
  \bibfield  {author} {\bibinfo {author} {\bibfnamefont {P.}~\bibnamefont
  {Basler}}\ and\ \bibinfo {author} {\bibfnamefont {M.}~\bibnamefont
  {M\"uhlleitner}},\ }\href {\doibase 10.1016/j.cpc.2018.11.006} {\bibfield
  {journal} {\bibinfo  {journal} {Comput. Phys. Commun.}\ }\textbf {\bibinfo
  {volume} {237}},\ \bibinfo {pages} {62} (\bibinfo {year} {2019})},\ \Eprint
  {http://arxiv.org/abs/1803.02846} {arXiv:1803.02846 [hep-ph]} \BibitemShut
  {NoStop}%
\bibitem [{\citenamefont {Ecker}\ \emph {et~al.}(1987)\citenamefont {Ecker},
  \citenamefont {Grimus},\ and\ \citenamefont {Neufeld}}]{Ecker:1987qp}%
  \BibitemOpen
  \bibfield  {author} {\bibinfo {author} {\bibfnamefont {G.}~\bibnamefont
  {Ecker}}, \bibinfo {author} {\bibfnamefont {W.}~\bibnamefont {Grimus}}, \
  and\ \bibinfo {author} {\bibfnamefont {H.}~\bibnamefont {Neufeld}},\ }\href
  {\doibase 10.1088/0305-4470/20/12/010} {\bibfield  {journal} {\bibinfo
  {journal} {J. Phys. A}\ }\textbf {\bibinfo {volume} {20}},\ \bibinfo {pages}
  {L807} (\bibinfo {year} {1987})}\BibitemShut {NoStop}%
\bibitem [{\citenamefont {Lavoura}\ and\ \citenamefont
  {Silva}(1994)}]{Lavoura:1994fv}%
  \BibitemOpen
  \bibfield  {author} {\bibinfo {author} {\bibfnamefont {L.}~\bibnamefont
  {Lavoura}}\ and\ \bibinfo {author} {\bibfnamefont {J.~P.}\ \bibnamefont
  {Silva}},\ }\href {\doibase 10.1103/PhysRevD.50.4619} {\bibfield  {journal}
  {\bibinfo  {journal} {Phys. Rev. D}\ }\textbf {\bibinfo {volume} {50}},\
  \bibinfo {pages} {4619} (\bibinfo {year} {1994})},\ \Eprint
  {http://arxiv.org/abs/hep-ph/9404276} {arXiv:hep-ph/9404276} \BibitemShut
  {NoStop}%
\bibitem [{\citenamefont {Ferreira}\ \emph {et~al.}(2009)\citenamefont
  {Ferreira}, \citenamefont {Haber},\ and\ \citenamefont
  {Silva}}]{Ferreira:2009wh}%
  \BibitemOpen
  \bibfield  {author} {\bibinfo {author} {\bibfnamefont {P.~M.}\ \bibnamefont
  {Ferreira}}, \bibinfo {author} {\bibfnamefont {H.~E.}\ \bibnamefont {Haber}},
  \ and\ \bibinfo {author} {\bibfnamefont {J.~P.}\ \bibnamefont {Silva}},\
  }\href {\doibase 10.1103/PhysRevD.79.116004} {\bibfield  {journal} {\bibinfo
  {journal} {Phys. Rev. D}\ }\textbf {\bibinfo {volume} {79}},\ \bibinfo
  {pages} {116004} (\bibinfo {year} {2009})},\ \Eprint
  {http://arxiv.org/abs/0902.1537} {arXiv:0902.1537 [hep-ph]} \BibitemShut
  {NoStop}%
\bibitem [{\citenamefont {Ferreira}\ \emph {et~al.}(2011)\citenamefont
  {Ferreira}, \citenamefont {Haber}, \citenamefont {Maniatis}, \citenamefont
  {Nachtmann},\ and\ \citenamefont {Silva}}]{Ferreira:2010yh}%
  \BibitemOpen
  \bibfield  {author} {\bibinfo {author} {\bibfnamefont {P.~M.}\ \bibnamefont
  {Ferreira}}, \bibinfo {author} {\bibfnamefont {H.~E.}\ \bibnamefont {Haber}},
  \bibinfo {author} {\bibfnamefont {M.}~\bibnamefont {Maniatis}}, \bibinfo
  {author} {\bibfnamefont {O.}~\bibnamefont {Nachtmann}}, \ and\ \bibinfo
  {author} {\bibfnamefont {J.~P.}\ \bibnamefont {Silva}},\ }\href {\doibase
  10.1142/S0217751X11051494} {\bibfield  {journal} {\bibinfo  {journal} {Int.
  J. Mod. Phys. A}\ }\textbf {\bibinfo {volume} {26}},\ \bibinfo {pages} {769}
  (\bibinfo {year} {2011})},\ \Eprint {http://arxiv.org/abs/1010.0935}
  {arXiv:1010.0935 [hep-ph]} \BibitemShut {NoStop}%
\bibitem [{\citenamefont {Ivanov}\ and\ \citenamefont
  {Silva}(2016)}]{Ivanov:2015mwl}%
  \BibitemOpen
  \bibfield  {author} {\bibinfo {author} {\bibfnamefont {I.~P.}\ \bibnamefont
  {Ivanov}}\ and\ \bibinfo {author} {\bibfnamefont {J.~P.}\ \bibnamefont
  {Silva}},\ }\href {\doibase 10.1103/PhysRevD.93.095014} {\bibfield  {journal}
  {\bibinfo  {journal} {Phys. Rev. D}\ }\textbf {\bibinfo {volume} {93}},\
  \bibinfo {pages} {095014} (\bibinfo {year} {2016})},\ \Eprint
  {http://arxiv.org/abs/1512.09276} {arXiv:1512.09276 [hep-ph]} \BibitemShut
  {NoStop}%
\bibitem [{\citenamefont {Ivanov}\ \emph
  {et~al.}(2019{\natexlab{a}})\citenamefont {Ivanov}, \citenamefont {Nishi},
  \citenamefont {Silva},\ and\ \citenamefont {Trautner}}]{Ivanov:2018ime}%
  \BibitemOpen
  \bibfield  {author} {\bibinfo {author} {\bibfnamefont {I.~P.}\ \bibnamefont
  {Ivanov}}, \bibinfo {author} {\bibfnamefont {C.~C.}\ \bibnamefont {Nishi}},
  \bibinfo {author} {\bibfnamefont {J.~a.~P.}\ \bibnamefont {Silva}}, \ and\
  \bibinfo {author} {\bibfnamefont {A.}~\bibnamefont {Trautner}},\ }\href
  {\doibase 10.1103/PhysRevD.99.015039} {\bibfield  {journal} {\bibinfo
  {journal} {Phys. Rev. D}\ }\textbf {\bibinfo {volume} {99}},\ \bibinfo
  {pages} {015039} (\bibinfo {year} {2019}{\natexlab{a}})},\ \Eprint
  {http://arxiv.org/abs/1810.13396} {arXiv:1810.13396 [hep-ph]} \BibitemShut
  {NoStop}%
\bibitem [{\citenamefont {Ivanov}(2017)}]{Ivanov:2017zjq}%
  \BibitemOpen
  \bibfield  {author} {\bibinfo {author} {\bibfnamefont {I.~P.}\ \bibnamefont
  {Ivanov}},\ }\href {\doibase 10.1088/1742-6596/873/1/012036} {\bibfield
  {journal} {\bibinfo  {journal} {J. Phys. Conf. Ser.}\ }\textbf {\bibinfo
  {volume} {873}},\ \bibinfo {pages} {012036} (\bibinfo {year} {2017})},\
  \Eprint {http://arxiv.org/abs/1702.07542} {arXiv:1702.07542 [hep-ph]}
  \BibitemShut {NoStop}%
\bibitem [{\citenamefont {Ivanov}\ and\ \citenamefont
  {Laletin}(2018)}]{Ivanov:2018qni}%
  \BibitemOpen
  \bibfield  {author} {\bibinfo {author} {\bibfnamefont {I.~P.}\ \bibnamefont
  {Ivanov}}\ and\ \bibinfo {author} {\bibfnamefont {M.}~\bibnamefont
  {Laletin}},\ }\href {\doibase 10.1103/PhysRevD.98.015021} {\bibfield
  {journal} {\bibinfo  {journal} {Phys. Rev. D}\ }\textbf {\bibinfo {volume}
  {98}},\ \bibinfo {pages} {015021} (\bibinfo {year} {2018})},\ \Eprint
  {http://arxiv.org/abs/1804.03083} {arXiv:1804.03083 [hep-ph]} \BibitemShut
  {NoStop}%
\bibitem [{\citenamefont {Ivanov}\ \emph
  {et~al.}(2019{\natexlab{b}})\citenamefont {Ivanov}, \citenamefont {Nishi},\
  and\ \citenamefont {Trautner}}]{Ivanov:2019kyh}%
  \BibitemOpen
  \bibfield  {author} {\bibinfo {author} {\bibfnamefont {I.~P.}\ \bibnamefont
  {Ivanov}}, \bibinfo {author} {\bibfnamefont {C.~C.}\ \bibnamefont {Nishi}}, \
  and\ \bibinfo {author} {\bibfnamefont {A.}~\bibnamefont {Trautner}},\ }\href
  {\doibase 10.1140/epjc/s10052-019-6845-x} {\bibfield  {journal} {\bibinfo
  {journal} {Eur. Phys. J. C}\ }\textbf {\bibinfo {volume} {79}},\ \bibinfo
  {pages} {315} (\bibinfo {year} {2019}{\natexlab{b}})},\ \Eprint
  {http://arxiv.org/abs/1901.11472} {arXiv:1901.11472 [hep-ph]} \BibitemShut
  {NoStop}%
\bibitem [{\citenamefont {Maniatis}\ \emph
  {et~al.}(2008{\natexlab{b}})\citenamefont {Maniatis}, \citenamefont {von
  Manteuffel},\ and\ \citenamefont {Nachtmann}}]{Maniatis:2007de}%
  \BibitemOpen
  \bibfield  {author} {\bibinfo {author} {\bibfnamefont {M.}~\bibnamefont
  {Maniatis}}, \bibinfo {author} {\bibfnamefont {A.}~\bibnamefont {von
  Manteuffel}}, \ and\ \bibinfo {author} {\bibfnamefont {O.}~\bibnamefont
  {Nachtmann}},\ }\href {\doibase 10.1140/epjc/s10052-008-0726-z} {\bibfield
  {journal} {\bibinfo  {journal} {Eur. Phys. J. C}\ }\textbf {\bibinfo {volume}
  {57}},\ \bibinfo {pages} {739} (\bibinfo {year} {2008}{\natexlab{b}})},\
  \Eprint {http://arxiv.org/abs/0711.3760} {arXiv:0711.3760 [hep-ph]}
  \BibitemShut {NoStop}%
\bibitem [{\citenamefont {Ferreira}\ and\ \citenamefont
  {Silva}(2011)}]{Ferreira:2010ir}%
  \BibitemOpen
  \bibfield  {author} {\bibinfo {author} {\bibfnamefont {P.~M.}\ \bibnamefont
  {Ferreira}}\ and\ \bibinfo {author} {\bibfnamefont {J.~P.}\ \bibnamefont
  {Silva}},\ }\href {\doibase 10.1103/PhysRevD.83.065026} {\bibfield  {journal}
  {\bibinfo  {journal} {Phys. Rev. D}\ }\textbf {\bibinfo {volume} {83}},\
  \bibinfo {pages} {065026} (\bibinfo {year} {2011})},\ \Eprint
  {http://arxiv.org/abs/1012.2874} {arXiv:1012.2874 [hep-ph]} \BibitemShut
  {NoStop}%
\bibitem [{\citenamefont {Ivanov}\ and\ \citenamefont
  {Nishi}(2013)}]{Ivanov:2013bka}%
  \BibitemOpen
  \bibfield  {author} {\bibinfo {author} {\bibfnamefont {I.~P.}\ \bibnamefont
  {Ivanov}}\ and\ \bibinfo {author} {\bibfnamefont {C.~C.}\ \bibnamefont
  {Nishi}},\ }\href {\doibase 10.1007/JHEP11(2013)069} {\bibfield  {journal}
  {\bibinfo  {journal} {JHEP}\ }\textbf {\bibinfo {volume} {11}},\ \bibinfo
  {pages} {069} (\bibinfo {year} {2013})},\ \Eprint
  {http://arxiv.org/abs/1309.3682} {arXiv:1309.3682 [hep-ph]} \BibitemShut
  {NoStop}%
\bibitem [{\citenamefont {Ivanov}(2009)}]{Ivanov:2008er}%
  \BibitemOpen
  \bibfield  {author} {\bibinfo {author} {\bibfnamefont {I.~P.}\ \bibnamefont
  {Ivanov}},\ }\href@noop {} {\bibfield  {journal} {\bibinfo  {journal} {Acta
  Phys. Polon. B}\ }\textbf {\bibinfo {volume} {40}},\ \bibinfo {pages} {2789}
  (\bibinfo {year} {2009})},\ \Eprint {http://arxiv.org/abs/0812.4984}
  {arXiv:0812.4984 [hep-ph]} \BibitemShut {NoStop}%
\bibitem [{\citenamefont {Ginzburg}\ \emph {et~al.}(2010)\citenamefont
  {Ginzburg}, \citenamefont {Ivanov},\ and\ \citenamefont
  {Kanishev}}]{Ginzburg:2009dp}%
  \BibitemOpen
  \bibfield  {author} {\bibinfo {author} {\bibfnamefont {I.~F.}\ \bibnamefont
  {Ginzburg}}, \bibinfo {author} {\bibfnamefont {I.~P.}\ \bibnamefont
  {Ivanov}}, \ and\ \bibinfo {author} {\bibfnamefont {K.~A.}\ \bibnamefont
  {Kanishev}},\ }\href {\doibase 10.1103/PhysRevD.81.085031} {\bibfield
  {journal} {\bibinfo  {journal} {Phys. Rev. D}\ }\textbf {\bibinfo {volume}
  {81}},\ \bibinfo {pages} {085031} (\bibinfo {year} {2010})},\ \Eprint
  {http://arxiv.org/abs/0911.2383} {arXiv:0911.2383 [hep-ph]} \BibitemShut
  {NoStop}%
\bibitem [{\citenamefont {Coleman}\ and\ \citenamefont
  {Weinberg}(1973)}]{Coleman:1973jx}%
  \BibitemOpen
  \bibfield  {author} {\bibinfo {author} {\bibfnamefont {S.~R.}\ \bibnamefont
  {Coleman}}\ and\ \bibinfo {author} {\bibfnamefont {E.~J.}\ \bibnamefont
  {Weinberg}},\ }\href {\doibase 10.1103/PhysRevD.7.1888} {\bibfield  {journal}
  {\bibinfo  {journal} {Phys. Rev. D}\ }\textbf {\bibinfo {volume} {7}},\
  \bibinfo {pages} {1888} (\bibinfo {year} {1973})}\BibitemShut {NoStop}%
\bibitem [{\citenamefont {Bernon}\ \emph {et~al.}(2018)\citenamefont {Bernon},
  \citenamefont {Bian},\ and\ \citenamefont {Jiang}}]{Bernon:2017jgv}%
  \BibitemOpen
  \bibfield  {author} {\bibinfo {author} {\bibfnamefont {J.}~\bibnamefont
  {Bernon}}, \bibinfo {author} {\bibfnamefont {L.}~\bibnamefont {Bian}}, \ and\
  \bibinfo {author} {\bibfnamefont {Y.}~\bibnamefont {Jiang}},\ }\href
  {\doibase 10.1007/JHEP05(2018)151} {\bibfield  {journal} {\bibinfo  {journal}
  {JHEP}\ }\textbf {\bibinfo {volume} {05}},\ \bibinfo {pages} {151} (\bibinfo
  {year} {2018})},\ \Eprint {http://arxiv.org/abs/1712.08430} {arXiv:1712.08430
  [hep-ph]} \BibitemShut {NoStop}%
\bibitem [{\citenamefont {Degee}\ and\ \citenamefont
  {Ivanov}(2010)}]{Degee:2009vp}%
  \BibitemOpen
  \bibfield  {author} {\bibinfo {author} {\bibfnamefont {A.}~\bibnamefont
  {Degee}}\ and\ \bibinfo {author} {\bibfnamefont {I.~P.}\ \bibnamefont
  {Ivanov}},\ }\href {\doibase 10.1103/PhysRevD.81.015012} {\bibfield
  {journal} {\bibinfo  {journal} {Phys. Rev. D}\ }\textbf {\bibinfo {volume}
  {81}},\ \bibinfo {pages} {015012} (\bibinfo {year} {2010})},\ \Eprint
  {http://arxiv.org/abs/0910.4492} {arXiv:0910.4492 [hep-ph]} \BibitemShut
  {NoStop}%
\bibitem [{\citenamefont {Quiros}(1999)}]{Quiros:1999jp}%
  \BibitemOpen
  \bibfield  {author} {\bibinfo {author} {\bibfnamefont {M.}~\bibnamefont
  {Quiros}},\ }in\ \href@noop {} {\emph {\bibinfo {booktitle} {{ICTP Summer
  School in High-Energy Physics and Cosmology}}}}\ (\bibinfo {year} {1999})\
  pp.\ \bibinfo {pages} {187--259},\ \Eprint
  {http://arxiv.org/abs/hep-ph/9901312} {arXiv:hep-ph/9901312} \BibitemShut
  {NoStop}%
\bibitem [{\citenamefont {Grzadkowski}\ \emph {et~al.}(2011)\citenamefont
  {Grzadkowski}, \citenamefont {Maniatis},\ and\ \citenamefont
  {Wudka}}]{Grzadkowski:2010dj}%
  \BibitemOpen
  \bibfield  {author} {\bibinfo {author} {\bibfnamefont {B.}~\bibnamefont
  {Grzadkowski}}, \bibinfo {author} {\bibfnamefont {M.}~\bibnamefont
  {Maniatis}}, \ and\ \bibinfo {author} {\bibfnamefont {J.}~\bibnamefont
  {Wudka}},\ }\href {\doibase 10.1007/JHEP11(2011)030} {\bibfield  {journal}
  {\bibinfo  {journal} {JHEP}\ }\textbf {\bibinfo {volume} {11}},\ \bibinfo
  {pages} {030} (\bibinfo {year} {2011})},\ \Eprint
  {http://arxiv.org/abs/1011.5228} {arXiv:1011.5228 [hep-ph]} \BibitemShut
  {NoStop}%
\bibitem [{\citenamefont {Zhitnitsky}(1980)}]{Zhitnitsky:1980tq}%
  \BibitemOpen
  \bibfield  {author} {\bibinfo {author} {\bibfnamefont {A.~R.}\ \bibnamefont
  {Zhitnitsky}},\ }\href@noop {} {\bibfield  {journal} {\bibinfo  {journal}
  {Sov. J. Nucl. Phys.}\ }\textbf {\bibinfo {volume} {31}},\ \bibinfo {pages}
  {260} (\bibinfo {year} {1980})}\BibitemShut {NoStop}%
\bibitem [{\citenamefont {Dine}\ \emph {et~al.}(1981)\citenamefont {Dine},
  \citenamefont {Fischler},\ and\ \citenamefont {Srednicki}}]{Dine:1981rt}%
  \BibitemOpen
  \bibfield  {author} {\bibinfo {author} {\bibfnamefont {M.}~\bibnamefont
  {Dine}}, \bibinfo {author} {\bibfnamefont {W.}~\bibnamefont {Fischler}}, \
  and\ \bibinfo {author} {\bibfnamefont {M.}~\bibnamefont {Srednicki}},\ }\href
  {\doibase 10.1016/0370-2693(81)90590-6} {\bibfield  {journal} {\bibinfo
  {journal} {Phys. Lett. B}\ }\textbf {\bibinfo {volume} {104}},\ \bibinfo
  {pages} {199} (\bibinfo {year} {1981})}\BibitemShut {NoStop}%
\bibitem [{\citenamefont {Mao}\ and\ \citenamefont {Zhu}(2014)}]{ZhuSCPV}%
  \BibitemOpen
  \bibfield  {author} {\bibinfo {author} {\bibfnamefont {Y.-n.}\ \bibnamefont
  {Mao}}\ and\ \bibinfo {author} {\bibfnamefont {S.-h.}\ \bibnamefont {Zhu}},\
  }\href {\doibase 10.1103/PhysRevD.90.115024} {\bibfield  {journal} {\bibinfo
  {journal} {Phys. Rev. D}\ }\textbf {\bibinfo {volume} {90}},\ \bibinfo
  {pages} {115024} (\bibinfo {year} {2014})},\ \Eprint
  {http://arxiv.org/abs/1409.6844} {arXiv:1409.6844 [hep-ph]} \BibitemShut
  {NoStop}%
\bibitem [{\citenamefont {Basler}\ \emph {et~al.}(2018)\citenamefont {Basler},
  \citenamefont {M\"uhlleitner},\ and\ \citenamefont
  {Wittbrodt}}]{Basler:2017uxn}%
  \BibitemOpen
  \bibfield  {author} {\bibinfo {author} {\bibfnamefont {P.}~\bibnamefont
  {Basler}}, \bibinfo {author} {\bibfnamefont {M.}~\bibnamefont
  {M\"uhlleitner}}, \ and\ \bibinfo {author} {\bibfnamefont {J.}~\bibnamefont
  {Wittbrodt}},\ }\href {\doibase 10.1007/JHEP03(2018)061} {\bibfield
  {journal} {\bibinfo  {journal} {JHEP}\ }\textbf {\bibinfo {volume} {03}},\
  \bibinfo {pages} {061} (\bibinfo {year} {2018})},\ \Eprint
  {http://arxiv.org/abs/1711.04097} {arXiv:1711.04097 [hep-ph]} \BibitemShut
  {NoStop}%
\bibitem [{\citenamefont {Ferreira}\ \emph {et~al.}(2023)\citenamefont
  {Ferreira}, \citenamefont {Grzadkowski}, \citenamefont {Ogreid},\ and\
  \citenamefont {Osland}}]{Ferreira:2023dke}%
  \BibitemOpen
  \bibfield  {author} {\bibinfo {author} {\bibfnamefont {P.~M.}\ \bibnamefont
  {Ferreira}}, \bibinfo {author} {\bibfnamefont {B.}~\bibnamefont
  {Grzadkowski}}, \bibinfo {author} {\bibfnamefont {O.~M.}\ \bibnamefont
  {Ogreid}}, \ and\ \bibinfo {author} {\bibfnamefont {P.}~\bibnamefont
  {Osland}},\ }\href@noop {} {\  (\bibinfo {year} {2023})},\ \Eprint
  {http://arxiv.org/abs/2306.02410} {arXiv:2306.02410 [hep-ph]} \BibitemShut
  {NoStop}%
\bibitem [{\citenamefont {Haber}\ \emph {et~al.}(1979)\citenamefont {Haber},
  \citenamefont {Kane},\ and\ \citenamefont {Sterling}}]{Haber:1978jt}%
  \BibitemOpen
  \bibfield  {author} {\bibinfo {author} {\bibfnamefont {H.~E.}\ \bibnamefont
  {Haber}}, \bibinfo {author} {\bibfnamefont {G.~L.}\ \bibnamefont {Kane}}, \
  and\ \bibinfo {author} {\bibfnamefont {T.}~\bibnamefont {Sterling}},\ }\href
  {\doibase 10.1016/0550-3213(79)90225-6} {\bibfield  {journal} {\bibinfo
  {journal} {Nucl. Phys. B}\ }\textbf {\bibinfo {volume} {161}},\ \bibinfo
  {pages} {493} (\bibinfo {year} {1979})}\BibitemShut {NoStop}%
\bibitem [{\citenamefont {Donoghue}\ and\ \citenamefont
  {Li}(1979)}]{Donoghue:1978cj}%
  \BibitemOpen
  \bibfield  {author} {\bibinfo {author} {\bibfnamefont {J.~F.}\ \bibnamefont
  {Donoghue}}\ and\ \bibinfo {author} {\bibfnamefont {L.~F.}\ \bibnamefont
  {Li}},\ }\href {\doibase 10.1103/PhysRevD.19.945} {\bibfield  {journal}
  {\bibinfo  {journal} {Phys. Rev. D}\ }\textbf {\bibinfo {volume} {19}},\
  \bibinfo {pages} {945} (\bibinfo {year} {1979})}\BibitemShut {NoStop}%
\bibitem [{\citenamefont {Hall}\ and\ \citenamefont
  {Wise}(1981)}]{Hall:1981bc}%
  \BibitemOpen
  \bibfield  {author} {\bibinfo {author} {\bibfnamefont {L.~J.}\ \bibnamefont
  {Hall}}\ and\ \bibinfo {author} {\bibfnamefont {M.~B.}\ \bibnamefont
  {Wise}},\ }\href {\doibase 10.1016/0550-3213(81)90469-7} {\bibfield
  {journal} {\bibinfo  {journal} {Nucl. Phys. B}\ }\textbf {\bibinfo {volume}
  {187}},\ \bibinfo {pages} {397} (\bibinfo {year} {1981})}\BibitemShut
  {NoStop}%
\bibitem [{\citenamefont {Barger}\ \emph {et~al.}(1990)\citenamefont {Barger},
  \citenamefont {Hewett},\ and\ \citenamefont {Phillips}}]{Barger:1989fj}%
  \BibitemOpen
  \bibfield  {author} {\bibinfo {author} {\bibfnamefont {V.~D.}\ \bibnamefont
  {Barger}}, \bibinfo {author} {\bibfnamefont {J.~L.}\ \bibnamefont {Hewett}},
  \ and\ \bibinfo {author} {\bibfnamefont {R.~J.~N.}\ \bibnamefont
  {Phillips}},\ }\href {\doibase 10.1103/PhysRevD.41.3421} {\bibfield
  {journal} {\bibinfo  {journal} {Phys. Rev. D}\ }\textbf {\bibinfo {volume}
  {41}},\ \bibinfo {pages} {3421} (\bibinfo {year} {1990})}\BibitemShut
  {NoStop}%
\bibitem [{\citenamefont {Barnett}\ \emph
  {et~al.}(1984{\natexlab{a}})\citenamefont {Barnett}, \citenamefont
  {Senjanovic}, \citenamefont {Wolfenstein},\ and\ \citenamefont
  {Wyler}}]{Barnett:1983mm}%
  \BibitemOpen
  \bibfield  {author} {\bibinfo {author} {\bibfnamefont {R.~M.}\ \bibnamefont
  {Barnett}}, \bibinfo {author} {\bibfnamefont {G.}~\bibnamefont {Senjanovic}},
  \bibinfo {author} {\bibfnamefont {L.}~\bibnamefont {Wolfenstein}}, \ and\
  \bibinfo {author} {\bibfnamefont {D.}~\bibnamefont {Wyler}},\ }\href
  {\doibase 10.1016/0370-2693(84)91179-1} {\bibfield  {journal} {\bibinfo
  {journal} {Phys. Lett. B}\ }\textbf {\bibinfo {volume} {136}},\ \bibinfo
  {pages} {191} (\bibinfo {year} {1984}{\natexlab{a}})}\BibitemShut {NoStop}%
\bibitem [{\citenamefont {Barnett}\ \emph
  {et~al.}(1984{\natexlab{b}})\citenamefont {Barnett}, \citenamefont
  {Senjanovic},\ and\ \citenamefont {Wyler}}]{Barnett:1984zy}%
  \BibitemOpen
  \bibfield  {author} {\bibinfo {author} {\bibfnamefont {R.~M.}\ \bibnamefont
  {Barnett}}, \bibinfo {author} {\bibfnamefont {G.}~\bibnamefont {Senjanovic}},
  \ and\ \bibinfo {author} {\bibfnamefont {D.}~\bibnamefont {Wyler}},\ }\href
  {\doibase 10.1103/PhysRevD.30.1529} {\bibfield  {journal} {\bibinfo
  {journal} {Phys. Rev. D}\ }\textbf {\bibinfo {volume} {30}},\ \bibinfo
  {pages} {1529} (\bibinfo {year} {1984}{\natexlab{b}})}\BibitemShut {NoStop}%
\bibitem [{\citenamefont {Jarlskog}(1985)}]{Jarlskog:1985ht}%
  \BibitemOpen
  \bibfield  {author} {\bibinfo {author} {\bibfnamefont {C.}~\bibnamefont
  {Jarlskog}},\ }\href {\doibase 10.1103/PhysRevLett.55.1039} {\bibfield
  {journal} {\bibinfo  {journal} {Phys. Rev. Lett.}\ }\textbf {\bibinfo
  {volume} {55}},\ \bibinfo {pages} {1039} (\bibinfo {year}
  {1985})}\BibitemShut {NoStop}%
\bibitem [{\citenamefont {Yu}\ and\ \citenamefont {Zhou}(2021)}]{Yu:2021cco}%
  \BibitemOpen
  \bibfield  {author} {\bibinfo {author} {\bibfnamefont {B.}~\bibnamefont
  {Yu}}\ and\ \bibinfo {author} {\bibfnamefont {S.}~\bibnamefont {Zhou}},\
  }\href {\doibase 10.1007/JHEP10(2021)017} {\bibfield  {journal} {\bibinfo
  {journal} {JHEP}\ }\textbf {\bibinfo {volume} {10}},\ \bibinfo {pages} {017}
  (\bibinfo {year} {2021})},\ \Eprint {http://arxiv.org/abs/2107.11928}
  {arXiv:2107.11928 [hep-ph]} \BibitemShut {NoStop}%
\end{thebibliography}%
\end{document}